\pdfoutput=1

\documentclass[12pt]{JHEP3}

\let\ifpdf\relax
\usepackage{ifpdf}
\usepackage{color}
\let\normalcolor\relax
\usepackage{mathtools}
\usepackage{cite}

\renewcommand{\hat}{\widehat}
\renewcommand{\tilde}{\widetilde}
\newcommand{\eq}[1]{\begin{equation}#1\end{equation}}
\newcommand{\ab}[1]{\langle #1\rangle}
\newcommand{\x}[2]{(#1,#2)}
\newcommand{\newcap}{\mathrm{\raisebox{0.75pt}{{$\,\bigcap\,$}}}}
\newcommand{\mi}{{\rm\rule[2.4pt]{6pt}{0.65pt}}}
\newcommand{\pl}{\hspace{0.5pt}\text{{\small+}}\hspace{-0.5pt}}
\newcommand{\ahat}{\hat{a}}
\definecolor{paper_blue}{rgb}{0.3,0.2,0.75}
\definecolor{hteal}{rgb}{0.0,0.425,0.1051}
\definecolor{perm}{rgb}{0.1,0.45,0.85}
\definecolor{varcolor}{rgb}{0.1,0.55,0.25}
\definecolor{functioncolor}{rgb}{0.1,0.35,0.75}
\definecolor{cut2}{rgb}{0.2353,0.2353,0.60}
\definecolor{cut1}{rgb}{0.60,0.0353,0.2353}
\newcommand{\mathematica}[3]{\vspace{0.35cm}\noindent\boxed{\begin{minipage}{#1\textwidth}\begin{tabular}{lp{11cm}}{\color{paper_blue}{\scriptsize{\tt In[1]:}}\raisebox{-0.65pt}{{\scriptsize{\tt=}}}}&{\tt #2}\\{\color{paper_blue}{\scriptsize {\tt Out[1]:}}\raisebox{-0.65pt}{{\scriptsize{\tt=}}}}&{\tt #3}\end{tabular}\end{minipage}}\vspace{0.35cm}}
\newcommand{\vardef}[1]{{\color{varcolor}{\sl #1}\rule[-1.05pt]{7.5pt}{.75pt}}}
\newcommand{\vardefms}[1]{{\color{varcolor}{\sl #1}\rule[-1.05pt]{15pt}{.75pt}}}
\newcommand{\vardefo}[1]{{\color{varcolor}{\sl #1}\rule[-1.05pt]{7.5pt}{.75pt}{\bf{\sl :}}}}

\newcommand{\defn}[3]{~\\[-35pt]\begin{itemize}\item[]\indent\hspace{-21pt}$\bullet$\hspace{-.75pt} {\tt {\color{functioncolor}#1}\![}#2{\tt\,]\!:}#3\end{itemize}\vspace{-10pt}}
\newcommand{\defnNA}[3]{~\\[-35pt]\begin{itemize}\item[]\indent\hspace{-21pt}$\bullet$\hspace{-.75pt} {\tt {\color{functioncolor}#1}\!}#2{\tt\,\!:}#3\end{itemize}\vspace{-10pt}}
\newcommand{\defntb}[4]{~\\[-35pt]\begin{itemize}\item[]\indent\hspace{-21pt}$\bullet$\hspace{-.75pt} {\tt {\color{functioncolor}#1}\![}#2{\tt\,]\![}#3{\tt\,]\!:}#4\end{itemize}\vspace{-10pt}}
\newcommand{\var}[1]{{\tt{\color{varcolor}{\sl#1}}}}
\newcommand{\ind}{\hspace{4ex}}
\newcommand{\fun}[1]{{\color{functioncolor}#1}}
\newcommand{\uscore}{\rule[-1.05pt]{7.5pt}{.75pt}}

\newcommand{\Li}{{\rm Li}_2}
\newcommand{\II}{\mathcal{I}}

\newcommand{\signA}{-}\newcommand{\signB}{+}\newcommand{\signC}{-}\newcommand{\signD}{}

\title{\mbox{\hspace{-0.35cm}{\LARGE Dual-Conformal Regularization of Infrared Loop}}\\
\mbox{\hspace{-0.35cm}{\LARGE Divergences and the {\it Chiral} Box Expansion}}}
\author{Jacob Bourjaily$^{a}$, Simon Caron-Huot$^{b,c}$, and Jaroslav Trnka$^{b,d}$\\
{\it $^{a}$ Department of Physics, Harvard University, Cambridge, MA 02138}\\
{\it $^{b}$ School of Natural Sciences, Institute for Advanced Study, Princeton, NJ 08540}\\
{\it $^{c}$ The Niels Bohr Institute, Copenhagen, Denmark DK-2100}\\
{\it $^{d}$ Department of Physics, Princeton University, Princeton, NJ 08544}}
\preprint{2013}
\abstract{
We revisit the familiar construction of one-loop scattering amplitudes via generalized unitarity in light of the recently understood properties of loop {\it integrands} prior to their integration. We show how in any four-dimensional quantum field theory, the {\it integrand}-level factorization of infrared divergences leads to twice as many constraints on integral coefficients than are visible from the integrated expressions.  In the case of planar, maximally supersymmetric Yang-Mills amplitudes, we demonstrate that these constraints are both sufficient and necessary to imply the finiteness and dual-conformal invariance of the ratios of scattering amplitudes. We present a novel regularization of the scalar box integrals which makes dual-conformal invariance of finite observables manifest term by term, and describe how this procedure can be generalized to higher loop-orders. Finally, we describe how the familiar scalar boxes at one-loop can be upgraded to `chiral boxes' resulting in a manifestly infrared-factorized, box-like expansion for all one-loop {\it integrands} in planar, $\mathcal{N}\!=\!4$ super Yang-Mills. Accompanying this note is a {\sc Mathematica} package which implements our results, and allows for the efficient numerical evaluation of any one-loop amplitude or ratio function.}
\preprint{PUPT-2441}

\begin{document}

\newpage

\section{Introduction}\label{introduction_section}
One-loop amplitudes have been extensively studied in recent decades, leading to many important insights and discoveries about the structure of scattering amplitudes, and frequently serve as an important source of theoretical `data' with which to test new ideas \cite{Bern:1993qk,Bern:1994zx,Bern:1994cg,Bern:1995ix,Britto:2004nj,Britto:2004nc,Drummond:2007cf,Cachazo:2008vp,CaronHuot:2010zt,Drummond:2010mb,ArkaniHamed:2012nw}. A powerful approach to computing loop amplitudes in any quantum field theory is the unitarity-based method, in which the amplitude is expanded into a basis of standardized scalar Feynman integrals (regulated if necessary) with coefficients fixed by on-shell scattering processes. Although very familiar and reasonably well understood, the way this approach has been realized in terms of existing technology does not make manifest several recently discovered aspects of loop-amplitudes---especially for the particularly rich case of scattering amplitudes in planar, $\mathcal{N}\!=\!4$ super Yang-Mills (SYM).

The two principle shortcomings about the way generalized unitarity has been realized in terms of existing tools (at least for $\mathcal{N}\!=\!4$ SYM) are: (1) that it fails to reflect the rich symmetries observed in loop amplitudes {\it prior to integration}; and (2) that even those symmetries which survive integration---such as the dual-conformal invariance (DCI) of the ratios of scattering amplitudes (see e.g.\ \cite{Drummond:2008vq})---are {\it severely} obfuscated in all existing regularization schema for infrared-divergent contributions. Because of this, manifestly-DCI expressions for ratio functions are known only in a few exceptionally simple cases \mbox{(see e.g.\ \cite{Drummond:2008vq,Drummond:2008bq,Elvang:2009ya,ArkaniHamed:2010kv,ArkaniHamed:2010gh})}. In this note, we revisit this story and fully address both shortcomings, providing manifestly-DCI expressions for {\it all} one-loop ratio functions in $\mathcal{N}\!=\!4$ SYM, and describing how the familiar box expansion can be upgraded to a {\it chiral} box expansion which matches all one-loop {\it integrands}.

This paper is organized as follows. In \mbox{section \ref{generalized_unitarity_section}} we review how generalized unitarity can be used to reproduce integrated amplitudes in $\mathcal{N}\!=\!4$ SYM, and in \mbox{section \ref{DCI_regulator_subsection}} we (heuristically) derive `DCI'-regularized expressions for all scalar box integrals, which are given in \mbox{Table \ref{dci_regulated_box_integrals}}. In \mbox{section \ref{leading_singularity_section}} we summarize the computation of scalar box coefficients using momentum-twistor variables, and write explicit formulae for all one-loop box coefficients in \mbox{Table \ref{one_loop_leading_singularities_table}}.

In \mbox{section \ref{general_discussion_of_DCI_regulator_section}} we explore the general features of the `DCI'-regularization proposal. In \mbox{section \ref{proof_of_dci_regularization_scheme_section}} we show that this proposal correctly reproduces all finite observables of any planar theory, thereby justifying its description as a `regularization scheme'. In \mbox{section \ref{higher_loop_regularization_section}} we describe how this scheme can be extended beyond one-loop, and compare it with existing approaches. The `DCI'-regulator is closely related to (and motivated by) the way that infrared divergences arise at the level of the loop-integrand. In \mbox{section \ref{cancelation_of_IR_divergences_section}}, we describe how the IR-divergences of loop amplitudes appear in terms of the `DCI'-regularization scheme, and in \mbox{section \ref{IR_equations_section}} we show how generalized unitarity realized at the {\it integrand}-level can be used to generate more powerful identities than would be possible after integration. In \mbox{section \ref{non_planar_generalization_section}} we illustrate the how these features persist beyond the planar-level.

In \mbox{section \ref{chiral_box_expansion_section}} we return to the familiar box expansion, and describe how it can be made `chiral' in a natural way, allowing us to match the full, chiral {\it integrand} of any one-loop scattering amplitude in $\mathcal{N}\!=\!4$.

For the purposes of concreteness and completeness, we review the basic kinematical variables---momentum twistors---used in most of this paper in \mbox{Appendix \ref{momentum_twistor_notation_and_review}}; and in \mbox{Appendix \ref{bcfw_loop_recursion_section}}, we use these variables to give a closed-form specialization of the BCFW recursion relations for all one-loop integrands in $\mathcal{N}\!=\!4$ SYM.

And finally, we have implemented the results described in this paper in a {\sc Mathematica} package called `{\tt loop\uscore amplitudes}' which is documented in \mbox{Appendix \ref{mathematica_implementation_section}}.

\newpage
\section{Revisiting Generalized Unitarity at One-Loop}\label{generalized_unitarity_section}

A major triumph of the unitarity-based approach to quantum field theory was the discovery that any one-loop amplitude can be written as a linear combination of standardized, scalar integrals with coefficients expressed as {\it on-shell diagrams}, historically known as `leading singularities' (for a comprehensive review, see \cite{ArkaniHamed:2012nw}). Because of the good UV-behavior of $\mathcal{N}\!=\!4$ SYM, only {\it box} integrals contribute---those involving {\it four} loop-momentum propagators---giving rise to the familiar `box expansion':
\vspace{-.2cm}\eq{\int\!\! d^4\ell\,\,\mathcal{A}_n^{(k),1}=\sum_{\!\!\!1\leq a<b<c<d\!\!\!}f_{a,b,c,d}\, I_{a,b,c,d},\vspace{-.2cm}\label{scalar_box_expansion}}
where $f_{a,b,c,d}$ are on-shell functions, and $I_{a,b,c,d}$ are the standard scalar box integrals,\\[-6pt]
\vspace{-.2cm}\eq{ f_{a,b,c,d}\equiv\raisebox{-50.pt}{\includegraphics[scale=1]{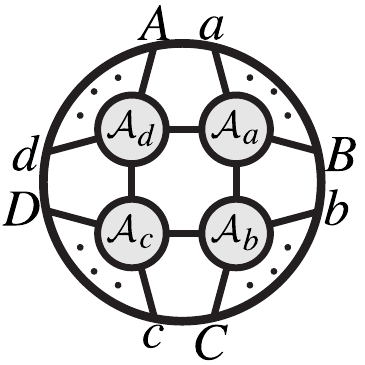}}\qquad I_{a,b,c,d}\equiv\raisebox{-50pt}{\includegraphics[scale=1]{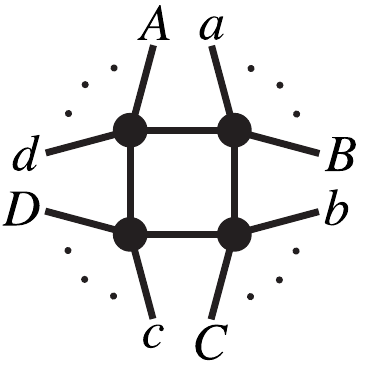}}\vspace{-.2cm}\label{terms_in_box_expansion}}
(To be clear, throughout this paper we will refer to $(\int\!\!d^4\ell)\mathcal{A}_n^{(k),1}$ as the (integrated) one-loop N$^{k}$MHV amplitude, expressed in units of $g^2N_c/(16\pi^2)$.)

The objects appearing in (\ref{scalar_box_expansion}) will each be described in detail below. But let us briefly remark on the motivation underlying the box expansion. Loop amplitudes are obtained by integrating the loop integrand over a four-dimensional contour of real loop momenta, $\ell\!\in\!\mathbb{R}^{3,1}$. If this integrand were obtained from the Feynman expansion, for example, it would include many propagators for the internal, `loop' particles. A co-dimension one residue of this integral `enclosing' one propagator---a `single-cut'---would correspond to putting one internal particle on-shell. Because the loop integral is four-dimensional, the highest-degree residues would be co-dimension four; these are the so-called {\it leading singularities} or `quad-cuts' of the integrand\footnote{In a general quantum field theory, the integrand may also have contributions where only co-dimension three (triangles) or co-dimension two (bubbles) residues exist; it is a non-trivial fact that $\mathcal{N}\!=\!4$ does not need contributions from triangles or bubbles (closely related to its UV-finiteness).} and are computed in terms of on-shell diagrams of the form shown in (\ref{terms_in_box_expansion}).

The scalar box integrals $I_{a,b,c,d}$ are simply those involving precisely four Feynman propagators of a scalar field theory---normalized to have co-dimension four residues of unit magnitude. As such, we should be able to represent any integrated loop amplitude in $\mathcal{N}\!=\!4$ by dressing each box integral with the actual co-dimension four residues `enclosing' the corresponding propagators as for the full loop integrand. A slight subtlety is that the residues of scalar boxes come in parity conjugate pairs, so in order to agree with the complete {\it integrand} the scalar boxes should be supplemented by parity-odd integrals, \cite{Cachazo:2008vp}.  Since these integrate to zero, they are often ignored. For the purpose of computing the {\it integrated} amplitude in this section---as opposed to the \emph{integrand}---we will also ignore them here.  This mismatch will be addressed in greater detail in \mbox{section \ref{chiral_box_expansion_section}}, where we show how to upgrade the box expansion (\ref{scalar_box_expansion}) in a way which allows us to match the full amplitude {\it prior to} integration.

\subsection{Scalar Box Integrals and their Divergences}\label{scalar_box_integrals_subsection}
Let us start our analysis with the generic, `four-mass', scalar box integral \cite{Bern:1992em,Bern:1993kr}---one for which all four corners are `massive' (involving at least two massless momenta):\\[-6pt]
\vspace{-.2cm}\eq{I_{a,b,c,d}\equiv\raisebox{-50pt}{\includegraphics[scale=1]{scalar_four_mass}}\equiv\int\!\!d^4\ell\,\,\frac{-\x{a}{c}\x{b}{d}\Delta\phantom{-}}{\x{\ell}{a}\x{\ell}{b}\x{\ell}{c}\x{\ell}{d}}\equiv\int\!\!d^4\ell\,\,\mathcal{I}_{a,b,c,d}(\ell),\vspace{-.2cm} \label{def_four_mass_box}}
where
\vspace{-.2cm}\eq{\Delta\equiv \sqrt{(1-u-v)^2-4 u v},\qquad u\equiv\frac{\x{a}{b}\x{c}{d}}{\x{a}{c}\x{b}{d}},\qquad v\equiv\frac{\x{b}{c}\x{a}{d}}{\x{a}{c}\x{b}{d}},\vspace{-.2cm}\label{definition_of_four_mass_delta}}
with $p_a\!\equiv\!x_{a+1}\,\mi\,x_a$ \cite{Drummond:2006rz} and where $1/\x{\ell}{a}$ denotes the standard propagator,
\vspace{-0.1cm}\eq{\x{a}{b}\equiv (x_a\,\mi\,x_b)^2 = (p_a\pl\,p_{a+1}\pl\,\cdots\pl\,p_{b-1})^2 \quad\mathrm{and}\quad \x{\ell}{a} \equiv(\ell\,\mi\, x_a)^2\,.\vspace{-0.1cm}}

When all the corners are massive, this integral is a transcendentality-two, completely finite, symmetric function of the dual-conformally invariant cross-ratios $(u,v)$:\\[-2pt]
\vspace{-.2cm}\eq{\boxed{\signA I_{a,b,c,d}(u,v)\equiv \signD\Li(\alpha)\signB\Li(\beta)\signC\Li(1)\signB\frac{1}{2}\log(u)\log(v)\signC\log(\alpha)\log(\beta),}\vspace{-.cm}\label{symmetric_four_mass_integral}}
with
\vspace{-.0cm}\eq{\alpha\equiv\frac{1}{2}(1-u+v+\Delta)\quad\mathrm{and}\quad \beta\equiv\frac{1}{2}(1+u-v+\Delta).\vspace{-.0cm}}
Notice that this form is {\it manifestly} symmetric under the exchange $u\!\leftrightarrow\!v$ as this exchange results only in $\alpha\!\leftrightarrow\!\beta$, under which (\ref{symmetric_four_mass_integral}) is obviously symmetric. The equivalence of (\ref{symmetric_four_mass_integral}) to existing formulae in the literature is easily verified\footnote{For notational simplicity, we have left implicit a factor $\pi^2$ in the definition of the measure of the integrand (\ref{def_four_mass_box})---the integrated expression, (\ref{symmetric_four_mass_integral}), is of course the standard one; we also omit factors of $i$ associated with the Wick rotation: ``$d^4\ell$'' denotes the Euclidean integration measure.}.

It is easy to see that this integral becomes divergent when any corner becomes massless---for example, identifying legs $p_a$ and $p_B$ results in:
\vspace{-.35cm}\eq{\raisebox{-50pt}{\includegraphics[scale=1]{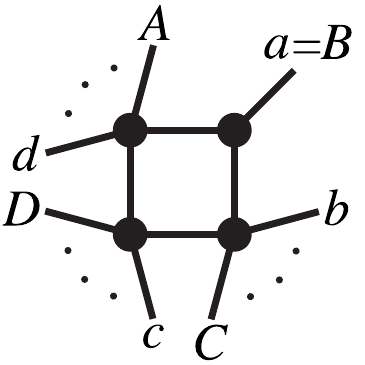}}\vspace{-.35cm}\label{unregulated_scalar_three_mass}}
This causes the cross ratio $u$ to vanish, introducing a logarithmic singularity from the term $-\frac{1}{2}\log(u)\log(v)$ in (\ref{symmetric_four_mass_integral}).
It is worth noting that (\ref{symmetric_four_mass_integral}) simplifies considerably when $u$ is taken to be parametrically small, $u\!\to\!\mathcal{O}(\epsilon)$: $\Delta\!\to\! (1\,\mi\,v)\pl\,\mathcal{O}(\epsilon)$, \mbox{$\alpha\!\to\!1\pl\,\mathcal{O}(\epsilon)$}, and $\beta\!\to\!(1\,\mi\,v)\pl\,\mathcal{O}(\epsilon)$, resulting in
\vspace{-.2cm}\eq{\signA I_{a,b,c,d}(u,v)\!\underset{u\to\mathcal{O}(\epsilon)}{=}\!\signD\Li(1-v)\signB\frac{1}{2}\log(u)\log(v)+\mathcal{O}(\epsilon).\vspace{-.2cm} \label{limit_of_four_mass_integral}}
This divergence can be regulated in a number of ways, including dimensional regularization (see e.g.\ \cite{Bern:1993kr} for standard formulae). Another canonical way to regulate such divergences uses the Higgs mechanism, \cite{Alday:2009zm}. In the simplest setup, this is used to give masses only to \emph{internal} propagators, and leads to the mass-regularized formulae found in e.g.\ \cite{Hodges:2010kq,Mason:2010pg}. However, because such regularization schema are predicated on a dimensionful parameter, the regulated formulae that result {\it severely} break dual-conformal invariance, obscuring the ultimate invariance of even absolutely convergent (and hence DCI) combinations of box integrals. (A complete basis involving only {\it manifestly} finite integrals for all convergent one-loop integrals was given in ref.\ \cite{ArkaniHamed:2010gh}.)

We are therefore motivated to find some way to make all the singular limits of the general integral (\ref{symmetric_four_mass_integral}) as dual-conformally invariant as possible, regulating the divergence caused by $u\!\to\!0$ in a way which depends only on some dimensionless parameter, denoted $\epsilon$, and dual-conformal cross-ratios. Such a regularization scheme is described in the next subsection.

\subsection{The `DCI' Regularization of Scalar Box Integrals}\label{DCI_regulator_subsection}

We propose the following regulator for one-loop integrals: render all external legs slightly off-shell by displacing the coordinates according to,
\vspace{-0.2cm}\eq{x_a\!\to\!\hat{x}_a\equiv x_a\pl\, \epsilon (x_{a+1}\,\mi\,x_a)\frac{\x{a\,\mi\,2}{a}}{\x{a\,\mi\,2}{a\pl1}}; \label{xshift}\vspace{-0.2cm}}
this transformation can be understood graphically as follows:
\vspace{-0.2cm}\eq{\raisebox{-70pt}{\includegraphics[scale=1]{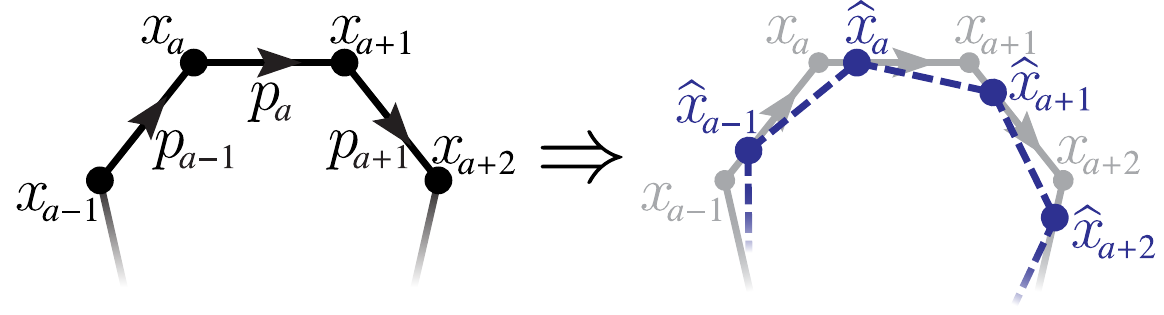}}\vspace{-1.0cm}\nonumber}
We will prove that this regulator produces the correct result for all finite observables and  discuss its implications, generalizations, and extensions in greater detail in \mbox{section \ref{general_discussion_of_DCI_regulator_section}}; but first, let us understand its consequences for the one-loop integrals appearing in the box expansion (\ref{scalar_box_expansion}).

If the dimensionless parameter $\epsilon$ is small (the only regime in which we will be interested) then the invariants $\x{a}{b}$ which are already non-vanishing are modified by a negligible amount.
However, when \mbox{$b\!=\!a\pl 1$} as in (\ref{unregulated_scalar_three_mass}), for example, the invariants will be regularized by (\ref{xshift}):
\vspace{-0.2cm}\eq{\x{a}{\hat b} = \epsilon \frac{\x{a\,\mi\,1}{b}\x{a}{b\pl1}}{\x{a\,\mi\,1}{b\pl1}} + \mathcal{O}(\epsilon^2)\quad \mathrm{with}\quad b\!=\!a\pl1\,.\vspace{-0.2cm}}
This implies that the integrated expressions for all box integrals will be given by limits of the four-mass-box (\ref{symmetric_four_mass_integral}); in particular, it is not necessary for us to add any `discontinuity functions' such as those discussed in e.g.\ \cite{Bern:1995ix}.

Consider the leading, degenerate limit of the box function, the so-called `three-mass' integral. This corresponds to taking $b\!=\!a\pl1$ in (\ref{symmetric_four_mass_integral}), keeping all other external corners massive: $\x{b}{c},\x{c}{d},\x{d}{a}\neq 0$.  The cross-ratio $v$
is unaffected by the regulator to $\mathcal{O}(\epsilon)$, while $u$ transforms according to
\vspace{-0.1cm}\eq{u \to \frac{\x{a}{\hat b}\x{c}{d}}{\x{a}{c}\x{b}{d}} = \epsilon \frac{\x{a\,\mi\,1}{b}\x{a}{b\pl1}\x{c}{d}}{\x{a\,\mi\,1}{b\pl1}\x{a}{c}\x{b}{d}}+\mathcal{O}(\epsilon^2)\equiv \epsilon u'+\mathcal{O}(\epsilon^2)\,.\vspace{-0.1cm}}
Using (\ref{limit_of_four_mass_integral}), the limit is easily seen to be given by
\vspace{-.3cm}\eq{\raisebox{-50pt}{\includegraphics[scale=1]{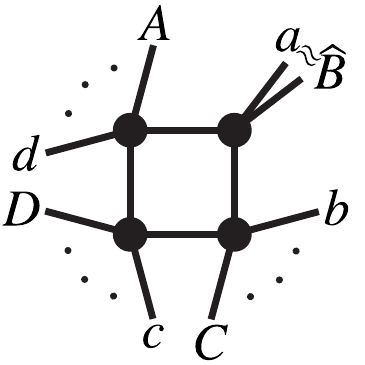}}=-\Li(1-v)-\frac{1}{2}\log(u')\log(v)-\frac{1}{2}\log(\epsilon)\log(v)+\mathcal{O}(\epsilon). \vspace{-.3cm}\label{regulated_three_mass}}

All further singular limits of (\ref{symmetric_four_mass_integral}) are similarly regulated according to the shifts (\ref{xshift}). This results in `DCI-regularized' forms for all scalar box integrals, which we have listed in detail in \mbox{Table \ref{dci_regulated_box_integrals}}. Importantly, these integrals depend {\it only} on the dimensionless parameter $\epsilon$ and dual-conformally invariant cross-ratios.

One special case not given in \mbox{Table\ref{dci_regulated_box_integrals}} is the so-called `massless' scalar box---which is only relevant for $n\!=\!4$. In this case, the shifts (\ref{xshift}) are not strictly defined, but can be generalized in simple way so that $u,v\!\to\!\epsilon^2$, resulting in the following, `DCI'-regulated massless box function:
\vspace{-0.2cm}\eq{\signA I^{\epsilon}_{1,2,3,4}=\signD\Li(1)\signB2\log(\epsilon)^2+\mathcal{O}(\epsilon)\qquad(\text{only for }n=4).}

\vspace{\fill}\newpage
\begin{table}[h*]\caption{{\bf `DCI'-Regulated Scalar Box Integrals}\label{dci_regulated_box_integrals}}
\vspace{-.3cm}\noindent\scalebox{.875}{\mbox{\hspace{1.07cm}\begin{minipage}[h]{\textwidth}\eq{\hspace{-2.5pt}\hspace{-2.0cm}\begin{array}{|@{}c@{}|@{}l@{}|}\hline\begin{array}{c}\raisebox{0pt}{\includegraphics[scale=1]{scalar_four_mass}}\\[-20pt]\rule{110pt}{.0pt}\\ \end{array}&\begin{array}{c@{}}\multicolumn{1}{l}{\rule{428pt}{.0pt}}\\[0.0pt]\multicolumn{1}{@{}l}{\text{{\large$\displaystyle\;\;\,\signA I=\signD\Li(\alpha)\signB\Li(\beta)\signC\Li(1)\signB\frac{1}{2}\log(u)\log(v)\signC\log(\alpha)\log(\beta)$}}}\\[20pt]\hline\multicolumn{1}{l@{}}{\!\text{{\normalsize$\begin{array}{@{}l@{}c}\\[-32pt]\multicolumn{1}{c}{}\!&\rule{050pt}{.0pt}\\\!&\begin{array}{lllllllllll@{}}\rule[-22.5pt]{.0pt}{50pt}\;\!u&\equiv&\displaystyle\frac{\x{a}{b}\x{c}{d}}{\x{a}{c}\x{b}{d}},\quad&v&\equiv&\displaystyle\frac{\x{b}{c}\x{a}{d}}{\x{a}{c}\x{b}{d}},\quad&\mathrm{and}&\quad\begin{array}{lcl}\alpha&\equiv&\frac{1}{2}(1\,\mi\,\,u\,\pl\,v\,\pl\,\Delta)\\\beta&\equiv&\frac{1}{2}(1\,\pl\,u\,\,\mi\,v\,\pl\,\Delta)\\\Delta&\equiv&\sqrt{(1\,\mi\,u\,\mi\,v)^2\,\mi\,4uv}\end{array}\end{array}\end{array}$}}}
\end{array}\\\hline\hline
\begin{array}{c}\raisebox{0pt}{\includegraphics[scale=1]{scalar_three_mass}}\\[-20pt]\rule{110pt}{.0pt}\\\hline  B=a\end{array}&\begin{array}{@{}c}\multicolumn{1}{@{}l}{\rule{428pt}{.0pt}}\\[15.0pt]\multicolumn{1}{@{}l}{\text{{\large$\displaystyle\;\;\,\signA I^{\epsilon}=\signD\Li(1-v)\signB\frac{1}{2}\log(u')\log(v)\signB\frac{1}{2}\log(\epsilon)\log(v)+\mathcal{O}(\epsilon)$}}}\\[35pt]\hline\multicolumn{1}{l@{}}{\!\text{{\normalsize$\begin{array}{@{}l@{}|c}\\[-32pt]\multicolumn{1}{c}{}&\rule{050pt}{.0pt}\\\begin{array}{@{}lll@{$\;$}}\;\;u&\mapsto&\epsilon\,u'\phantom{{}^2}\\\hline \;\;v&\mapsto&v\end{array}&\begin{array}{llllllll@{}}\rule[-14.5pt]{.0pt}{35pt}u'&\equiv&\displaystyle\frac{\x{a\,\mi\,1}{b}\x{a}{b\pl1}\x{c}{d}}{\x{a\,\mi\,1}{b\pl1}\x{a}{c}\x{b}{d}},&\quad &\hspace{40pt}v&\equiv&\displaystyle\frac{\x{b}{c}\x{a}{d}}{\x{a}{c}\x{b}{d}}\end{array}\end{array}$}}}
\end{array}\\\hline\hline
\begin{array}{c}\raisebox{0pt}{\includegraphics[scale=1]{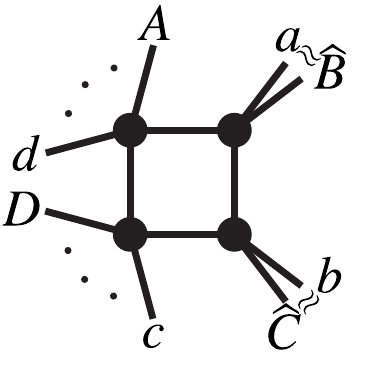}}\\[-20pt]\rule{110pt}{.0pt}\\\hline  B=a,\quad C=b\end{array}&\begin{array}{c@{}}\multicolumn{1}{l}{\rule{428pt}{.0pt}}\\[15.0pt]\multicolumn{1}{@{}l}{\text{{\large$\displaystyle\,\;\;\signA I^{\epsilon}=\signD\Li(1)\signB\frac{1}{2}\log(u')\log(v')\signB\frac{1}{2}\log(\epsilon)\log(u'v')\signB\frac{1}{2}\log(\epsilon)^2+\mathcal{O}(\epsilon)$}}}\\[35pt]\hline\multicolumn{1}{l@{}}{\!\text{{\normalsize$\begin{array}{@{}l@{}|c}\\[-32pt]\multicolumn{1}{c}{}&\rule{050pt}{.0pt}\\\begin{array}{@{}lll@{$\;$}}\;\;u&\mapsto&\epsilon\,u'\phantom{{}^2}\\\hline \;\;v&\mapsto&\epsilon\,v'\end{array}&\begin{array}{llllll@{}}\rule[-14.5pt]{.0pt}{35pt}u'&\equiv&\displaystyle\frac{\x{a\,\mi\,1}{b}\x{c}{d}}{\x{a\,\mi\,1}{c}\x{b}{d}},\quad\qquad&\hspace{7pt}\hspace{41.5pt}\quad v'&\equiv&\displaystyle\frac{\x{b}{c\pl1}\x{a}{d}}{\x{b}{d}\x{a}{c\pl1}}\end{array}\end{array}$}}}
\end{array}\\\hline\hline
\begin{array}{c}\raisebox{0pt}{\includegraphics[scale=1]{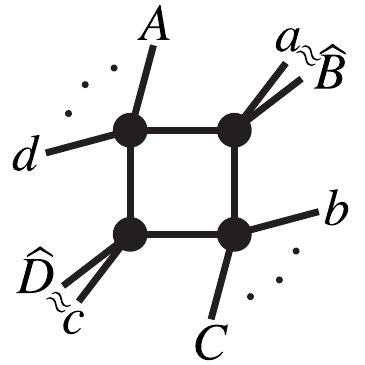}}\\[-20pt]\rule{110pt}{.0pt}\\\hline  B=a,\quad D=c\end{array}&\begin{array}{c@{}}\multicolumn{1}{l}{\rule{428pt}{.0pt}}\\[15.0pt]\multicolumn{1}{@{}l}{\text{{\large$\displaystyle\,\;\;\signA I^{\epsilon}=\signD\Li(1-v)\signB\frac{1}{2}\log(u')\log(v)\signB\log(\epsilon)\log(v)+\mathcal{O}(\epsilon)$}}}\\[35pt]\hline\multicolumn{1}{l@{}}{\!\text{{\normalsize$\begin{array}{@{}l@{}|c}\\[-32pt]\multicolumn{1}{c}{}&\rule{050pt}{.0pt}\\\begin{array}{@{}lll@{$\;$}}\;\;u&\mapsto&\epsilon^2\,u'\\\hline \;\;v&\mapsto&v\end{array}&\begin{array}{llllll@{}}\rule[-14.5pt]{.0pt}{35pt}u'&\equiv&\displaystyle\frac{\x{a\,\mi\,1}{b}\x{a}{b\pl1}\x{c\,\mi\,1}{d}\x{c}{d\pl1}}{\x{a\,\mi\,1}{b\pl1}\x{c\,\mi\,1}{d\pl1}\x{a}{c}\x{b}{d}},&v&\equiv&\displaystyle\frac{\x{b}{c}\x{a}{d}}{\x{a}{c}\x{b}{d}}\end{array}\end{array}$}}}
\end{array}\\\hline\hline
\begin{array}{c}\raisebox{0pt}{\includegraphics[scale=1]{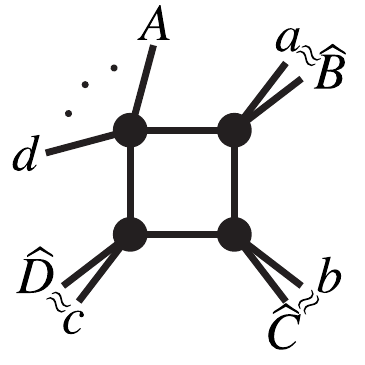}}\\[-20pt]\rule{110pt}{.0pt}\\\hline  B=a,\; C=b,\;D=c\end{array}&\begin{array}{c@{}}\multicolumn{1}{l}{\rule{428pt}{.0pt}}\\[15.0pt]\multicolumn{1}{@{}l}{\text{{\large$\displaystyle\,\;\;\signA I^{\epsilon}=\signD\Li(1)\signB\frac{1}{2}\log(\epsilon)\log(u')\signB\log(\epsilon)^2+\mathcal{O}(\epsilon)$}}}\\[35pt]\hline\multicolumn{1}{l@{}}{\!\text{{\normalsize$\begin{array}{@{}l@{}|c}\\[-32pt]\multicolumn{1}{c}{}&\rule{050pt}{.0pt}\\\begin{array}{@{}lll@{$\;$}}\;\;u&\mapsto&\epsilon^2\,u'\\\hline \;\;v&\mapsto&\epsilon\!\times\!1\end{array}&\begin{array}{llllll@{}}\rule[-14.5pt]{.0pt}{35pt}u'&\equiv&\displaystyle\frac{\x{a\,\mi\,1}{b}\x{c}{d\pl1}}{\x{a\,\mi\,1}{c}\x{b}{d\pl1}}&\end{array}\end{array}$}}}
\end{array}\\\hline
\end{array}\nonumber}\end{minipage}}}
\end{table}
\vspace{\fill}\newpage

\subsection{Scalar Box Coefficients: One-Loop Leading Singularities}\label{leading_singularity_section}
Factorization dictates that the residues of the loop integrand---called {\it leading singularities} $\!\!$---are simply the products of tree-amplitudes, summed-over all the internal particles which can be exchanged, and integrated over the on-shell phase space of each. As mentioned above, we represent such functions graphically as {\it on-shell diagrams} of the form shown in (\ref{terms_in_box_expansion}).

These are simply algebraic (super)functions of the external kinematical data---almost always rational, and at one-loop involving at most the solution to a quadratic equation. Leading singularities have of course been known to the literature for quite some time, and can be computed in many ways. A comprehensive summary of these objects---their classification, evaluation, and relations---was described recently in \mbox{ref.\ \cite{ArkaniHamed:2012nw}.}
The physical content of any on-shell diagram (after blowing up all tree-amplitudes at the vertices themselves into on-shell diagrams) is encoded by a permutation \cite{ArkaniHamed:2012nw}, and the permutations of the corners are simply `glued' together to give the permutation of the `one-loop' diagram:
\vspace{-.1cm}\eq{\raisebox{-51pt}{\includegraphics[scale=.85]{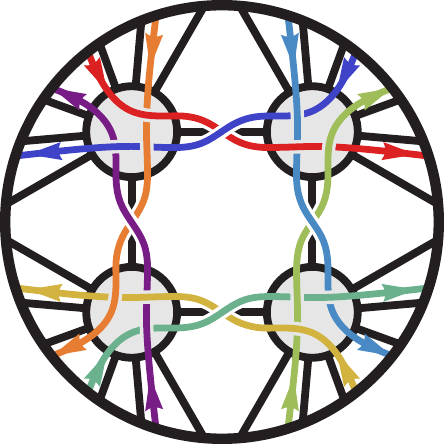}}\vspace{-.1cm}}
Given the permutation labeling an on-shell diagram, it is trivial to construct an explicit formula for the corresponding on-shell function. This is done most directly in terms of an auxiliary Grassmannian integral as described in \mbox{ref.\ \cite{ArkaniHamed:2012nw}}. (All the necessary tools involved in this story have been made available in a {\sc Mathematica} package called `{\tt positroids}', which is documented in \mbox{ref.\ \cite{Bourjaily:2012gy}}.) We will not review these ideas here, but simply give the formulae that result.

The most compact expressions for leading singularities are found when they are written in terms of the momentum-twistor variables introduced in ref.\ \cite{Hodges:2009hk} since they simultaneously trivialize the two ubiquitous kinematical constraints---the on-shell condition and momentum conservation. Momentum-twistors are simply the twistor-variables \cite{Penrose:1967wn} associated to the region-momentum coordinates $x_a$ defined in \mbox{section \ref{generalized_unitarity_section}}. (A brief introduction to momentum-twistor variables and an explanation of the notation used throughout this section is given in \mbox{Appendix \ref{momentum_twistor_notation_and_review}}.)

Assuming a modicum of familiarity with momentum-twistors, let us now describe the form that leading singularities take. We start with the most general case: a leading singularity involving four massive corners. It turns out that this case is the only one we need to consider, as it will smoothly generate all the others in a very natural way.

The most important data are the two solutions ${\color{cut1}\ell_1},{\color{cut2}\ell_2}$ to the kinematical constraints of putting all the internal lines on-shell,
\vspace{-.3cm}\eq{\raisebox{-62pt}{\includegraphics[scale=0.9]{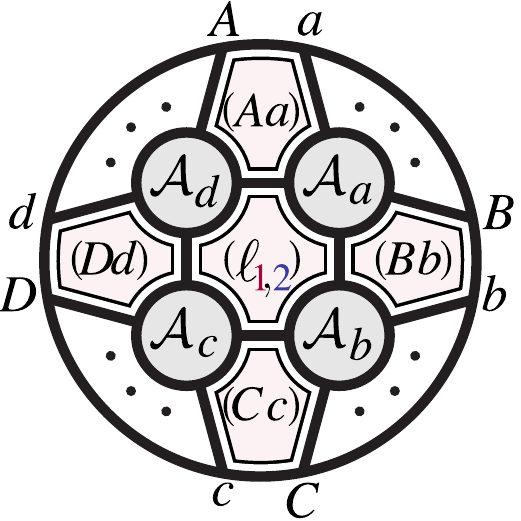}}\vspace{-.3cm}}
Here, the lines $(Aa),\ldots,(Dd)$ in momentum-twistor space correspond to the region-momenta $x_a,\ldots,x_d$ used in \mbox{section \ref{scalar_box_integrals_subsection}}; and the lines ${\color{cut1}\ell_{1}}$ and ${\color{cut2}\ell_{2}}$ correspond to the two solutions to the problem of putting four propagators on-shell. These `quad-cuts' are the solution to a simple geometry problem in momentum-twistor space (viewed projectively as $\mathbb{P}^3$):  ${\color{cut1}\ell_{1}}$ and ${\color{cut2}\ell_{2}}$ are the two lines which simultaneously intersect the four generic lines  $(Aa),\ldots,(Dd)$:\\[-8pt]
\vspace{-.1cm}\eq{\raisebox{-62pt}{\includegraphics[scale=0.9]{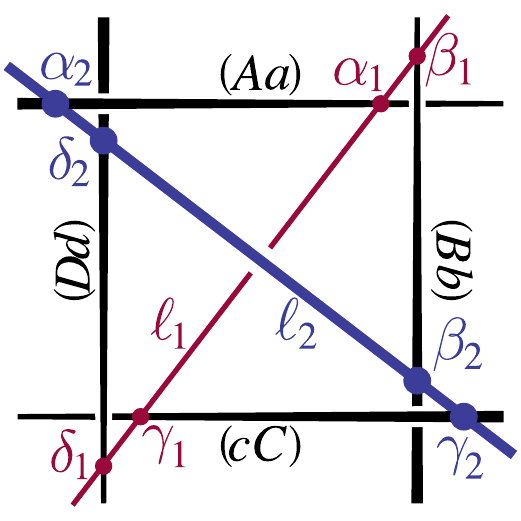}}\vspace{-.1cm}\label{four_mass_quad_cut_figure}}
(The fact that there are {\it two} solutions to the problem of putting four-propagators on-shell is a classic result of the Schubert calculus---and continues to hold even when the four lines are non-generic; see \mbox{ref.\ \cite{ArkaniHamed:2010gh}} for an exposition of these ideas.)

We will give explicit formulae for the twistors ${\color{cut1}\alpha_1},\ldots,{\color{cut1}\delta_1}$ and ${\color{cut2}\alpha_2},\ldots,{\color{cut2}\delta_2}$ corresponding to the two lines ${\color{cut1}\ell_{1}}$ and ${\color{cut2}\ell_{2}}$ intersecting the lines $(Aa),\ldots,(Dd)$, respectively, in \mbox{Table \ref{quad_cuts_table}}; but for now let us take it for granted that they are known. Given these twistors, it is easy to write the leading singularity for each particular quad-cut $\ell_{{\color{cut1}1},{\color{cut2}2}}$ in terms of momentum-twistors.  In momentum-twistor space, we are dealing with MHV-stripped amplitudes (so (N$^{k=0}$)MHV tree-amplitudes are simply the identity), and polarization sums become simple multiplication of the corresponding MHV-stripped amplitudes. This allows us to `peel-off' the tree-amplitudes at each corner from a standard on-shell graph involving only MHV amplitudes, \cite{Drummond:2008bq}:
\vspace{-.25cm}\eq{\raisebox{-61.6pt}{\includegraphics[scale=.875]{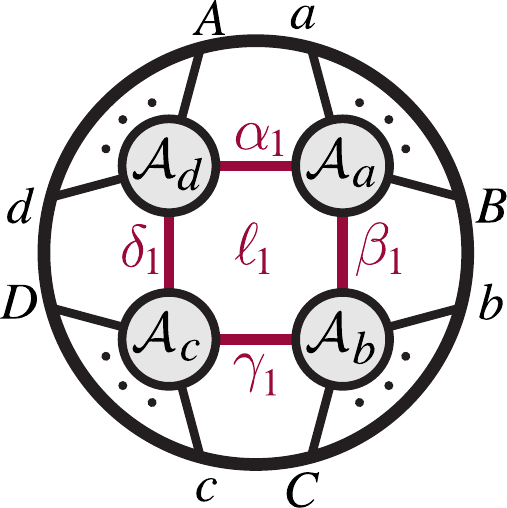}}=\left(\!\!\begin{array}{@{}c@{}l@{}}&\mathcal{A}_a({\color{cut1}\alpha_1},a,\ldots,B,{\color{cut1}\beta_1})\\
\times&\mathcal{A}_b({\color{cut1}\beta_1}\,,b,\ldots,C,{\color{cut1}\gamma_1\,})\\
\times&\mathcal{A}_c({\color{cut1}\gamma_1}\,,c,\ldots,D,{\color{cut1}\delta_1\,})\\
\times&\mathcal{A}_d({\color{cut1}\delta_1}\,,d,\ldots,A,{\color{cut1}\alpha_1})\end{array}\right)\!\times\raisebox{-61.6pt}{\includegraphics[scale=.875]{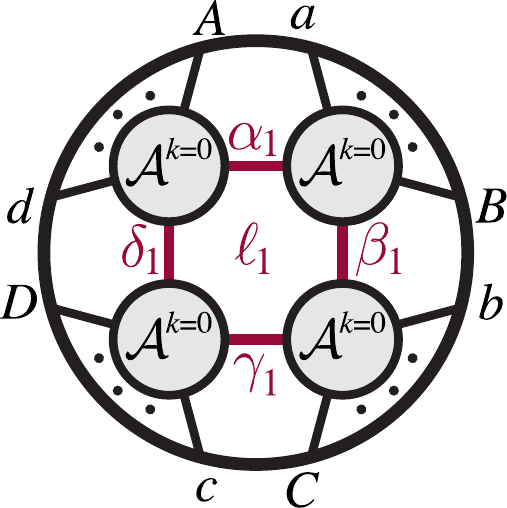}}
\vspace{-.25cm}\nonumber}
where the on-shell graph on the right is the N$^{k=2}$MHV `four-mass' function\footnote{We note that this  expression for the four-mass function appears to differ from that given in \mbox{ref.\ \cite{Mason:2009qx}} which we suspect to be incomplete.} \cite{ArkaniHamed:2012nw},
\vspace{-.25cm}\eq{\hspace{-4.15cm}\raisebox{-61.6pt}{\includegraphics[scale=.875]{four_mass_1_mhv_corners}}=\![{\color{cut1}\delta_1}\,A\,a\,B\,b][{\color{cut1}\beta_1}\,C\,c\,D\,d]\!\left(\!\!1-\frac{\ab{{\color{cut1}\beta_1}\,d\,Aa}\ab{{\color{cut1}\delta_1}\,b\,Cc}}{\ab{{\color{cut1}\beta_1}\,d\,Cc}\ab{{\color{cut1}\delta_1}\,b\,Aa}}\!\right)^{-1}\!\!\!\!\!\!\!.\hspace{-3cm}\vspace{-.15cm}\label{four_mass_coefficient_1}}
While simply replacing $({\color{cut1}\alpha_1},\ldots,{\color{cut1}\delta_1})\!\to\!({\color{cut2}\alpha_2},\ldots,{\color{cut2}\delta_2})$ in the formula above would give (minus) the other leading singularity, we will find it advantageous to use an alternate form of the four-mass function involving the ${\color{cut2}\ell_2}$ solution:
\vspace{-.15cm}\eq{\hspace{-3.975cm}\raisebox{-61.6pt}{\includegraphics[scale=.875]{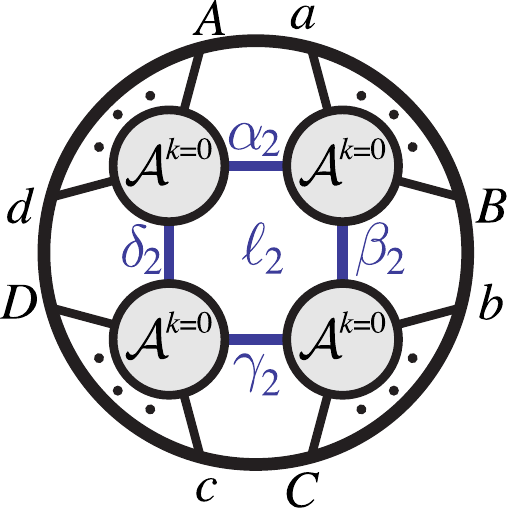}}=\![{\color{cut2}\alpha_2}\,B\,b\,C\,c][{\color{cut2}\gamma_2}\,D\,d\,A\,a]\!\left(\!\!1-\frac{\ab{{\color{cut2}\alpha_2}\,c\,Dd}\ab{{\color{cut2}\gamma_2}\,a\,Bb}}{\ab{{\color{cut2}\alpha_2}\,c\,Bb}\ab{{\color{cut2}\gamma_2}\,a\,Dd}}\!\right)^{-1}\!\!\!\!\!\!\!.\hspace{-3cm}\vspace{-.05cm}\label{four_mass_coefficient_2}}

In \mbox{Table \ref{quad_cuts_table}}, we give the particular solutions to the quad-cuts represented graphically in (\ref{four_mass_quad_cut_figure})---where $\Delta$ is defined as in equation (\ref{definition_of_four_mass_delta}). (The notation used here is fully defined in \mbox{Appendix\ \ref{momentum_twistor_notation_and_review}}.) The motivation for using two separate formulae for the four-mass leading singularities is that they separately evolve smoothly to all other cases. This is made possible by the fact that the multiplicative factors appearing in \mbox{Table \ref{one_loop_leading_singularities_table}}, which encode the shifts of the quad-cuts from each `corner' of the box (see equation (\ref{four_mass_quad_cut_figure})), are all smooth and non-singular in limits where some of the legs are identified. (Notice that $\varphi_{{\color{cut1}1},{\color{cut2}2}}\!\to\!1$ in all limits where a pair of legs are identified.)

As promised, the formulae given above for the four-mass leading singularities smoothly generate {\it all} one-loop leading singularities, which for the purposes of completeness and reference have been written explicitly in \mbox{Table \ref{one_loop_leading_singularities_table}}. The formulae given in \mbox{Table \ref{one_loop_leading_singularities_table}} have been organized in order to highlight how each case descends smoothly from the general cases given above for the four-mass functions.

The complete box coefficient for a given topology is given by the sum of the two corresponding on-shell diagrams, $f_{a,b,c,d}\!\equiv\!{\color{cut1}f^1_{a,b,c,d}}+{\color{cut2}f^2_{a,b,c,d}}$, which involve the quad-cuts ${\color{cut1}\ell_1}$ and ${\color{cut2}\ell_2}$, respectively:
\vspace{-.2cm}\eq{\text{{\large$\displaystyle{\color{cut1}f^1_{a,b,c,d}}\equiv$}{\LARGE$$}}\raisebox{-72pt}{\includegraphics[scale=1]{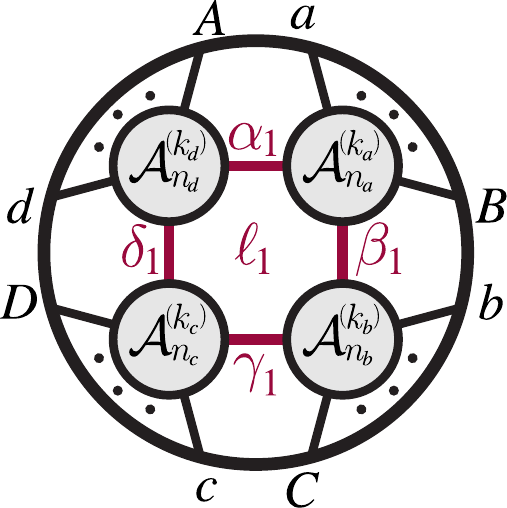}}\qquad\text{{\large$\displaystyle{\color{cut2}f^2_{a,b,c,d}}\equiv$}{\LARGE$$}}\raisebox{-72pt}{\includegraphics[scale=1]{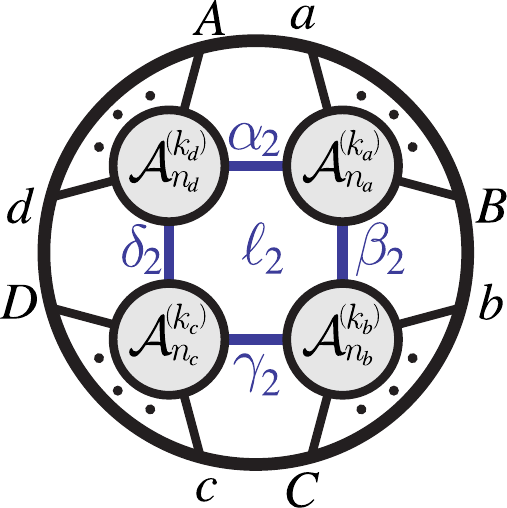}}\vspace{-.2cm}\nonumber}
Here, each graph represents a sum over all such graphs with the same topology---involving any four amplitudes $\mathcal{A}_{n_a}^{(k_a)},\ldots,\mathcal{A}_{n_d}^{(k_d)}$ such that $k_a\pl\,k_b\pl\,k_c\pl\,k_d=k\,\mi\,2$, and for which $n_a\pl\,n_b\pl\,n_c\pl\,n_d\!=\!n\pl8$, with $0\leq\!k\leq n\,\mi\,4$ for each (except when $n\!=\!3$, for which $\mathcal{A}_3^{(-1)}$ is allowed).

\begin{table}[b*]{\small\vspace{-1cm}\eq{\hspace{-3cm}\begin{array}{|l@{}l@{}l@{}c@{}l|}\hline&&&&\\[-12pt]
\text{{\normalsize${\color{cut1}\alpha_1}$}}&\equiv&(aA)\!\newcap\!(Bb{\color{cut1}\,\gamma_1})&\equiv&\text{{\normalsize$z_a\pl z_A$}}\displaystyle\frac{\ab{a\,Bb\,(cC)\!\newcap\!(Dd\,A)}\pl\,\ab{A\,Bb\,(cC)\!\newcap\!(Dd\,a)}\pl\,\ab{aA\,cC}\ab{Bb\,Dd}\Delta}{2\ab{Bb\,(cC)\!\newcap\!(Dd\,A)\,A}}\\[10pt]&&&&\\[-12pt]
\text{{\normalsize${\color{cut1}\beta_1}$}}&\equiv&(Bb)\!\newcap\!(aA{\color{cut1}\,\delta_1})&\equiv&\text{{\normalsize$z_B\pl z_b$}} \displaystyle\frac{\ab{B\,aA\,(Dd)\!\newcap\!(cC\,b)}\pl\,\ab{b\,aA\,(Dd)\!\newcap\!(cC\,B)}\pl\,\ab{aA\,cC}\ab{Bb\,Dd}\Delta}{2\ab{aA\,(Dd)\!\newcap\!(cC\,b)\,b}}\\[10pt]&&&&\\[-10pt]
\multicolumn{5}{|c|}{\text{{\normalsize${\color{cut1}\gamma_1}$}}\equiv(cC)\!\newcap\!(Dd{\color{cut1}\,\alpha_1})\quad\mathrm{and}\quad\text{{\normalsize${\color{cut1}\delta_1}$}}\equiv(Dd)\!\newcap\!(cC{\color{cut1}\,\beta_1})}\\
&&&&\\[-5pt]\hline\hline
\text{{\normalsize${\color{cut2}\alpha_2}$}}&\equiv&(Aa)\!\newcap\!(dD{\color{cut2}\,\gamma_2})&\equiv&\text{{\normalsize$z_A\pl z_a$}}\displaystyle\frac{\ab{A\,dD\,(Cc)\!\newcap\!(bB\,a)}\pl\,\ab{a\,dD\,(Cc)\!\newcap\!(bB\,A)}\pl\,\ab{Aa\,Cc}\ab{bB\,dD}\Delta}{2\ab{dD\,(Cc)\!\newcap\!(bB\,a)\,a}}\\[10pt]&&&&\\[-12pt]
\text{{\normalsize${\color{cut2}\beta_2}$}}&\equiv&(bB)\!\newcap\!(Cc{\color{cut2}\,\delta_2})&\equiv&\text{{\normalsize$z_b\pl z_B$}}\displaystyle\frac{\ab{b\,Cc\,(dD)\!\newcap\!(Aa\,B)}\pl\,\ab{B\,Cc\,(dD)\!\newcap\!(Aa\,b)}\pl\,\ab{Aa\,Cc}\ab{bB\,dD}\Delta}{2\ab{Cc\,(dD)\!\newcap\!(Aa\,B)\,B}}\\[10pt]&&&&\\[-10pt]
\multicolumn{5}{|c|}{\text{{\normalsize${\color{cut2}\gamma_2}$}}\equiv(Cc)\!\newcap\!(bB{\color{cut2}\,\alpha_2})\quad\mathrm{and}\quad\text{{\normalsize${\color{cut2}\delta_2}$}}\equiv(dD)\!\newcap\!(Aa{\color{cut2}\,\beta_2})}\\[-5pt]
&&&&\\[0pt]\hline
\end{array}\nonumber\hspace{-3cm}\vspace{-.6cm}}}
\caption{Explicit solutions ${\color{cut1}\ell_1},{\color{cut2}\ell_2}$ to the Schubert problem involving four generic lines.\label{quad_cuts_table}}\vspace{-1.5cm}
\end{table}

~\newpage
\begin{table}[h*]\caption{{\bf All One-Loop Leading Singularities in Momentum-Twistor Space}\label{one_loop_leading_singularities_table}}\vspace{-.3cm}
\noindent\scalebox{.91}{\mbox{\hspace{1.07cm}\begin{minipage}[h]{\textwidth}\vspace{0cm}\eq{\hspace{-3pt}\hspace{-1.95cm}\begin{array}{@{}c@{}}\begin{array}{|c|@{}c@{}|}\hline\raisebox{-85pt}{\includegraphics[scale=.75]{four_mass_1}}&
\rule[-22.5pt]{0pt}{50pt}\begin{array}{c}{\color{cut1}\varphi_1}[{\color{cut1}\delta_1}\,A\,a\,B\,b][{\color{cut1}\beta_1}\,C\,c\,D\,d]\times\\\mathcal{A}_{a}({\color{cut1}\alpha_1},\ldots,{\color{cut1}\beta_1})\mathcal{A}_b({\color{cut1}\beta_1},\ldots,{\color{cut1}\gamma_1})\\\mathcal{A}_c({\color{cut1}\gamma_1},\ldots,{\color{cut1}\delta_1})\,\mathcal{A}_d({\color{cut1}\delta_1},\ldots,{\color{cut1}\alpha_1})\end{array}
\\[-62pt]\cline{2-2}&\begin{array}{lll}\hspace{-0.5cm}{\color{cut1}\ell_1\raisebox{-1pt}{$\left\{\rule[-20pt]{0pt}{52pt}\right.$}}\hspace{-0.2cm}&\begin{array}{r@{}lr@{}l}&\rule{85pt}{0pt}\\[-16pt]{\color{cut1}\alpha_1}&=\!(aA)\!\newcap\!(Bb\,{\color{cut1}\gamma_1})\\{\color{cut1}\beta_1}&=\!(Bb)\!\newcap\!(aA\,{\color{cut1}\delta_1})
\\{\color{cut1}\gamma_1}&=\!(cC)\!\newcap\!(Dd\,{\color{cut1}\alpha_1})\\{\color{cut1}\delta_1}&=\!(Dd)\!\newcap\!(cC\,{\color{cut1}\beta_1})
\end{array}&\hspace{-0.3cm}{\color{cut1}\raisebox{-1pt}{$\left.\rule[-20pt]{0pt}{52pt}\right\}$}}\hspace{-0.2cm}\end{array}\\\hline
\raisebox{-85pt}{\includegraphics[scale=.75]{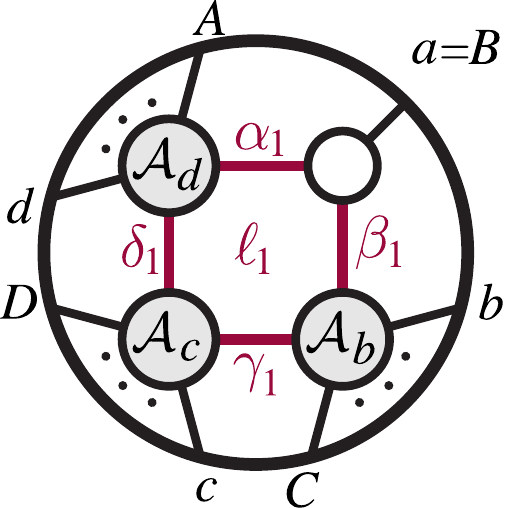}}&
\rule[-22.5pt]{0pt}{50pt}\begin{array}{c}[{\color{cut1}\beta_1}\,C\,c\,D\,d]\!\times\!\mathcal{A}_b({\color{cut1}\beta_1},\ldots,{\color{cut1}\gamma_1})\\\mathcal{A}_c({\color{cut1}\gamma_1},\ldots,{\color{cut1}\delta_1})\mathcal{A}_d({\color{cut1}\delta_1},\ldots,{\color{cut1}\alpha_1})\end{array}
\\[-62pt]\cline{2-2}&\begin{array}{lll}\hspace{-0.5cm}{\color{cut1}\ell_1\raisebox{-1pt}{$\left\{\rule[-20pt]{0pt}{52pt}\right.$}}\hspace{-0.2cm}&\begin{array}{r@{}lr@{}l}&\rule{85pt}{0pt}\\[-16pt]{\color{cut1}\alpha_1}&=\!a\\{\color{cut1}\beta_1}&=\!B
\\{\color{cut1}\gamma_1}&=\!(cC)\!\newcap\!(Dda)\\{\color{cut1}\delta_1}&=\!(Dd)\!\newcap\!(cCB)
\end{array}&\hspace{-0.3cm}{\color{cut1}\raisebox{-1pt}{$\left.\rule[-20pt]{0pt}{52pt}\right\}$}}\hspace{-0.2cm}\end{array}\\\hline
\raisebox{-85pt}{\includegraphics[scale=.75]{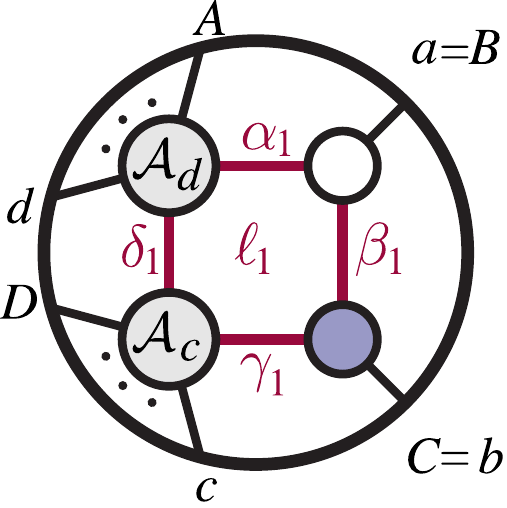}}&
\rule[-22.5pt]{0pt}{50pt}\begin{array}{c}\phantom{\varphi_1}[{\color{cut1}\beta_1}\,C\,c\,D\,d]\times\\\mathcal{A}_c({\color{cut1}\gamma_1},\ldots,{\color{cut1}\delta_1})\mathcal{A}_d({\color{cut1}\delta_1},\ldots,{\color{cut1}\alpha_1})\end{array}
\\[-62pt]\cline{2-2}&\begin{array}{lll}\hspace{-0.5cm}{\color{cut1}\ell_1\raisebox{-1pt}{$\left\{\rule[-20pt]{0pt}{52pt}\right.$}}\hspace{-0.2cm}&\begin{array}{r@{}lr@{}l}&\rule{85pt}{0pt}\\[-16pt]{\color{cut1}\alpha_1}&=\!a\\{\color{cut1}\beta_1}&=\!B
\\{\color{cut1}\gamma_1}&=\!(cC)\!\newcap\!(Dda)\\{\color{cut1}\delta_1}&=\!(Dd)\!\newcap\!(cCB)
\end{array}&\hspace{-0.3cm}{\color{cut1}\raisebox{-1pt}{$\left.\rule[-20pt]{0pt}{52pt}\right\}$}}\hspace{-0.2cm}\end{array}\\\hline
\raisebox{-85pt}{\includegraphics[scale=.75]{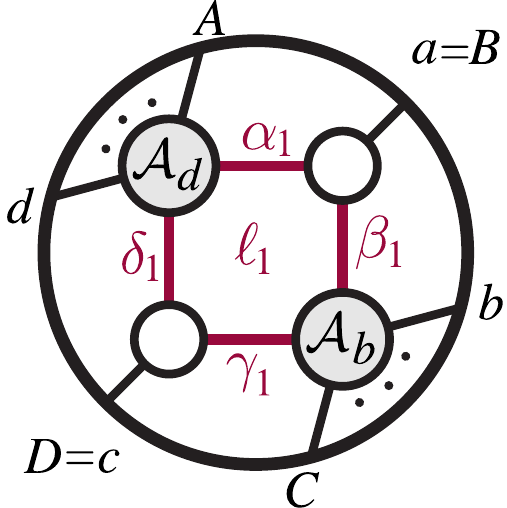}}&
\rule[-22.5pt]{0pt}{50pt}\begin{array}{c}\mathcal{A}_b({\color{cut1}\beta_1},\ldots,{\color{cut1}\gamma_1})\mathcal{A}_d({\color{cut1}\delta_1},\ldots,{\color{cut1}\alpha_1})\end{array}
\\[-62pt]\cline{2-2}&\begin{array}{lll}\hspace{-0.5cm}{\color{cut1}\ell_1\raisebox{-1pt}{$\left\{\rule[-20pt]{0pt}{52pt}\right.$}}\hspace{-0.2cm}&\begin{array}{r@{}lr@{}l}&\rule{85pt}{0pt}\\[-16pt]{\color{cut1}\alpha_1}&=\!a\\{\color{cut1}\beta_1}&=\!B
\\{\color{cut1}\gamma_1}&=\!c\\{\color{cut1}\delta_1}&=\!D
\end{array}&\hspace{-0.3cm}{\color{cut1}\raisebox{-1pt}{$\left.\rule[-20pt]{0pt}{52pt}\right\}$}}\hspace{-0.2cm}\end{array}\\\hline
\raisebox{-85pt}{\includegraphics[scale=.75]{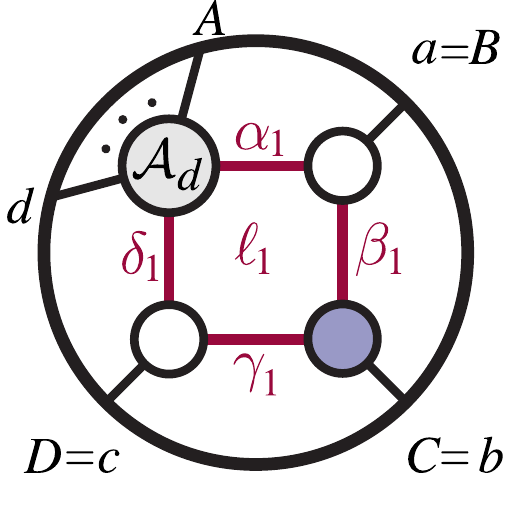}}&
\rule[-22.5pt]{0pt}{50pt}\begin{array}{c}\mathcal{A}_d({\color{cut1}\delta_1},\ldots,{\color{cut1}\alpha_1})\end{array}
\\[-62pt]\cline{2-2}&\begin{array}{lll}\hspace{-0.5cm}{\color{cut1}\ell_1\raisebox{-1pt}{$\left\{\rule[-20pt]{0pt}{52pt}\right.$}}\hspace{-0.2cm}&\begin{array}{r@{}lr@{}l}&\rule{85pt}{0pt}\\[-16pt]{\color{cut1}\alpha_1}&=\!a\\{\color{cut1}\beta_1}&=\!B
\\{\color{cut1}\gamma_1}&=\!c\\{\color{cut1}\delta_1}&=\!D
\end{array}&\hspace{-0.3cm}{\color{cut1}\raisebox{-1pt}{$\left.\rule[-20pt]{0pt}{52pt}\right\}$}}\hspace{-0.2cm}\end{array}\\\hline
\end{array}\,\begin{array}{|@{}c@{}|c|}\hline
\rule[-22.5pt]{0pt}{50pt}\begin{array}{c}{\color{cut2}\varphi_2}[{\color{cut2}\alpha_2}\,B\,b\,C\,c][{\color{cut2}\gamma_2}\,D\,d\,A\,a]\times\\\mathcal{A}_a({\color{cut2}\alpha_2},\ldots,{\color{cut2}\beta_2})\mathcal{A}_b({\color{cut2}\beta_2},\ldots,{\color{cut2}\gamma_2})\\\mathcal{A}_c({\color{cut2}\gamma_2},\ldots,{\color{cut2}\delta_2})\,\mathcal{A}_d({\color{cut2}\delta_2},\ldots,{\color{cut2}\alpha_2})\end{array}&\raisebox{-85pt}{\includegraphics[scale=.75]{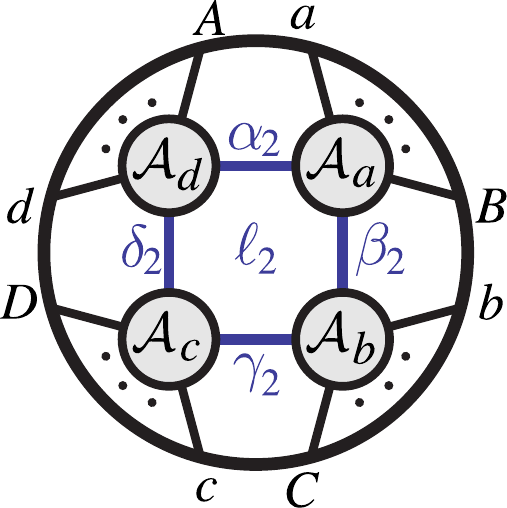}}
\\[-62pt]\cline{1-1}\begin{array}{lll}\hspace{-0.5cm}{\color{cut2}\ell_2\raisebox{-1pt}{$\left\{\rule[-20pt]{0pt}{52pt}\right.$}}\hspace{-0.2cm}&\begin{array}{r@{}lr@{}l}&\rule{85pt}{0pt}\\[-16pt]{\color{cut2}\alpha_2}&=\!(Aa)\!\newcap\!(dD\,{\color{cut2}\gamma_2})\\{\color{cut2}\beta_2}&=\!(bB)\!\newcap\!(Cc\,{\color{cut2}\delta_2})
\\{\color{cut2}\gamma_2}&=\!(Cc)\!\newcap\!(bB\,{\color{cut2}\alpha_2})\\{\color{cut2}\delta_2}&=\!(dD)\!\newcap\!(Aa\,{\color{cut2}\beta_2})
\end{array}&\hspace{-0.3cm}{\color{cut2}\raisebox{-1pt}{$\left.\rule[-20pt]{0pt}{52pt}\right\}$}}\hspace{-0.2cm}\end{array}&\\\hline
\rule[-22.5pt]{0pt}{50pt}\begin{array}{c}\phantom{\varphi_2}[{\color{cut2}\alpha_2}\,B\,b\,C\,c][{\color{cut2}\gamma_2}\,D\,d\,A\,a]\times\\\mathcal{A}_b({\color{cut2}\beta_2},\ldots,{\color{cut2}\gamma_2})\\\mathcal{A}_c({\color{cut2}\gamma_2},\ldots,{\color{cut2}\delta_2})\,\mathcal{A}_d({\color{cut2}\delta_2},\ldots,{\color{cut2}\alpha_2})\end{array}&\raisebox{-85pt}{\includegraphics[scale=.75]{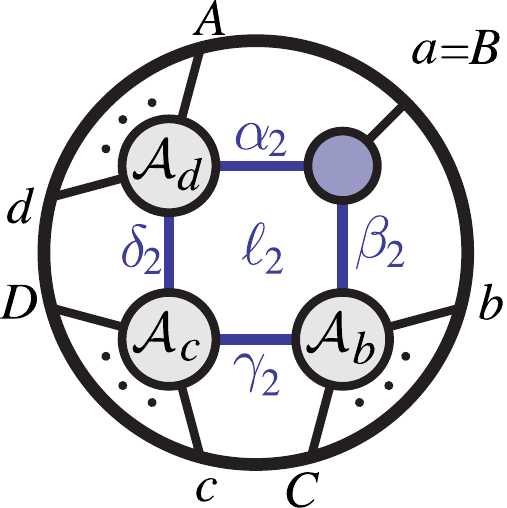}}
\\[-62pt]\cline{1-1}\begin{array}{lll}\hspace{-0.5cm}{\color{cut2}\ell_2\raisebox{-1pt}{$\left\{\rule[-20pt]{0pt}{52pt}\right.$}}\hspace{-0.2cm}&\begin{array}{r@{}lr@{}l}&\rule{85pt}{0pt}\\[-16pt]{\color{cut2}\alpha_2}&=\!(Aa)\!\newcap\!(dD\,{\color{cut2}\gamma_2})\\{\color{cut2}\beta_2}&=\!(bB)\!\newcap\!(Cc\,{\color{cut2}\delta_2})
\\{\color{cut2}\gamma_2}&=\!(Cc)\!\newcap\!(bBA)\\{\color{cut2}\delta_2}&=\!(dD)\!\newcap\!(Aab)
\end{array}&\hspace{-0.3cm}{\color{cut2}\raisebox{-1pt}{$\left.\rule[-20pt]{0pt}{52pt}\right\}$}}\hspace{-0.2cm}\end{array}&\\\hline
\rule[-22.5pt]{0pt}{50pt}\begin{array}{c}[{\color{cut2}\gamma_2}\,D\,d\,A\,a]\times\\\mathcal{A}_c({\color{cut2}\gamma_2},\ldots,{\color{cut2}\delta_2})\,\mathcal{A}_d({\color{cut2}\delta_2},\ldots,{\color{cut2}\alpha_2})\end{array}&\raisebox{-85pt}{\includegraphics[scale=.75]{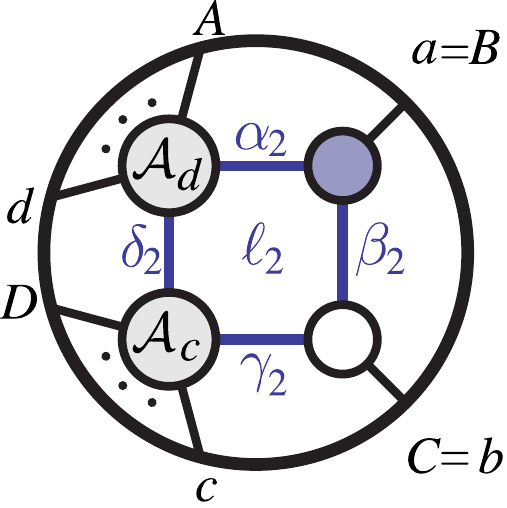}}
\\[-62pt]\cline{1-1}\begin{array}{lll}\hspace{-0.5cm}{\color{cut2}\ell_2\raisebox{-1pt}{$\left\{\rule[-20pt]{0pt}{52pt}\right.$}}\hspace{-0.2cm}&\begin{array}{r@{}lr@{}l}&\rule{85pt}{0pt}\\[-16pt]{\color{cut2}\alpha_2}&=\!(Aa)\!\newcap\!(dDC)\\{\color{cut2}\beta_2}&=\!b
\\{\color{cut2}\gamma_2}&=\!C\\{\color{cut2}\delta_2}&=\!(dD)\!\newcap\!(Aab)
\end{array}&\hspace{-0.3cm}{\color{cut2}\raisebox{-1pt}{$\left.\rule[-20pt]{0pt}{52pt}\right\}$}}\hspace{-0.2cm}\end{array}&\\\hline
\rule[-22.5pt]{0pt}{50pt}\begin{array}{c}\phantom{\varphi_2}[{\color{cut2}\alpha_2}\,B\,b\,C\,c][{\color{cut2}\gamma_2}\,D\,d\,A\,a]\times\\\mathcal{A}_b({\color{cut2}\beta_2},\ldots,{\color{cut2}\gamma_2})\mathcal{A}_d({\color{cut2}\delta_2},\ldots,{\color{cut2}\alpha_2})\end{array}&\raisebox{-85pt}{\includegraphics[scale=.75]{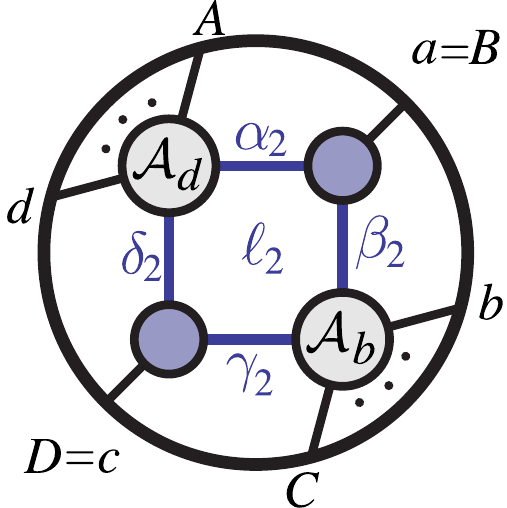}}
\\[-62pt]\cline{1-1}\begin{array}{lll}\hspace{-0.5cm}{\color{cut2}\ell_2\raisebox{-1pt}{$\left\{\rule[-20pt]{0pt}{52pt}\right.$}}\hspace{-0.2cm}&\begin{array}{r@{}lr@{}l}&\rule{85pt}{0pt}\\[-16pt]{\color{cut2}\alpha_2}&=\!(Aa)\!\newcap\!(dDC)\\{\color{cut2}\beta_2}&=\!(bB)\!\newcap\!(Ccd)
\\{\color{cut2}\gamma_2}&=\!(Cc)\!\newcap\!(bBA)\\{\color{cut2}\delta_2}&=\!(dD)\!\newcap\!(Aab)
\end{array}&\hspace{-0.3cm}{\color{cut2}\raisebox{-1pt}{$\left.\rule[-20pt]{0pt}{52pt}\right\}$}}\hspace{-0.2cm}\end{array}&\\\hline
\rule[-22.5pt]{0pt}{50pt}\begin{array}{c}[{\color{cut2}\gamma_2}\,D\,d\,A\,a]\!\times\!\mathcal{A}_d({\color{cut2}\delta_2},\ldots,{\color{cut2}\alpha_2})\end{array}&\raisebox{-85pt}{\includegraphics[scale=.75]{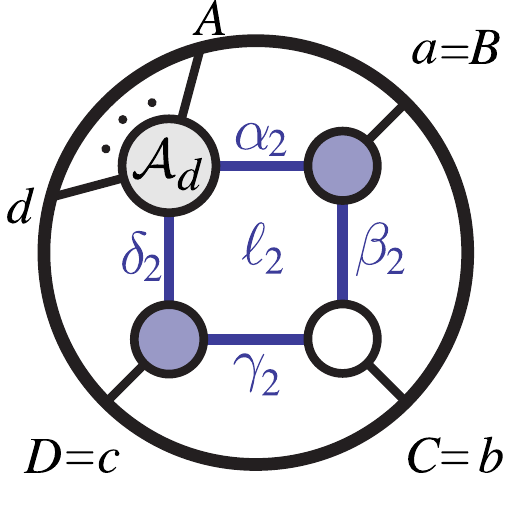}}
\\[-62pt]\cline{1-1}\begin{array}{lll}\hspace{-0.5cm}{\color{cut2}\ell_2\raisebox{-1pt}{$\left\{\rule[-20pt]{0pt}{52pt}\right.$}}\hspace{-0.2cm}&\begin{array}{r@{}lr@{}l}&\rule{85pt}{0pt}\\[-16pt]{\color{cut2}\alpha_2}&=\!(Aa)\!\newcap\!(dDC)\\{\color{cut2}\beta_2}&=\!b
\\{\color{cut2}\gamma_2}&=\!C\\{\color{cut2}\delta_2}&=\!(dD)\!\newcap\!(Aab)
\end{array}&\hspace{-0.3cm}{\color{cut2}\raisebox{-1pt}{$\left.\rule[-20pt]{0pt}{52pt}\right\}$}}\hspace{-0.2cm}\end{array}&\\\hline
\end{array}\end{array}
\nonumber\vspace{-.2cm}}
\eq{\begin{array}{c}\displaystyle\mathrm{with}\qquad{\color{cut1}\varphi_1}\equiv\left(1-\frac{\ab{{\color{cut1}\beta_1}\,d\,Aa}\ab{{\color{cut1}\delta_1}\,b\,Cc}}{\ab{{\color{cut1}\beta_1}\,d\,Cc}\ab{{\color{cut1}\delta_1}\,b\,Aa}}\right)^{-1}\!\!\!\!\quad\mathrm{and}\quad{\color{cut2}\varphi_2}\equiv\left(1-\frac{\ab{{\color{cut2}\alpha_2}\,c\,Dd}\ab{{\color{cut2}\gamma_2}\,a\,Bb}}{\ab{{\color{cut2}\alpha_2}\,c\,Bb}\ab{{\color{cut2}\gamma_2}\,a\,Dd}}\right)^{-1}\!\!\!\!.\,\,\,\,\qquad\phantom{\mathrm{with}}\\[15pt]\displaystyle
\hspace{-1.5cm}\text{(Cyclically-related leading singularities are related by }{\color{cut1}f^1_{a,b,c,d}}={\color{cut2}f^2_{b,c,d,a}}.)
\qquad\qquad~\end{array}\nonumber\vspace{-1.9cm}}\end{minipage}}}
\end{table}
\vspace{\fill}\newpage

\section{Properties and Extensions of `DCI' Regularization}\label{general_discussion_of_DCI_regulator_section}
Using the `DCI'-regularized box integrals $I^\epsilon_{a,b,c,d}$ and the box coefficients described in \mbox{section \ref{generalized_unitarity_section}}, the scalar-box expansion becomes
\vspace{-.1cm}\eq{\int\!\!d^4\ell\,\,\mathcal{A}_n^{(k),1}= \sum_{\!\!\!1\leq a<b<c<d\!\!\!}\Big({\color{cut1}f^1_{a,b,c,d}}+{\color{cut2}f^2_{a,b,c,d}}\Big)\,I^\epsilon_{a,b,c,d}.\vspace{-.2cm}\label{scalar_box_expansion_two_types_of_coefficients}}
Let us now prove that this produces the correct result for all finite observables in planar $\mathcal{N}\!=\!4$ SYM by showing that the expressions given for $I^{\epsilon}_{a,b,c,d}$ can be obtained from a very concrete and simple regularization procedure.

\subsection{The `DCI' Regularization Scheme}\label{proof_of_dci_regularization_scheme_section}

Given any four-dimensional integrand $\II(\ell)$, we define its `DCI'-regulated integral by deforming the integrand in the following way:
\vspace{-0.2cm}\eq{\int\limits_{\textrm{reg}}\!\!d^4\ell \,\,\II(\ell) \equiv \int\!\!d^4 \ell \,\,\II(\ell)R(\ell)\quad\mbox{where}\quad R(\ell)\equiv\prod_{a=1}^n \frac{\x{\ell}{a}}{\x{\ell}{\hat a}}\,. \label{definition_of_regulator}\vspace{-0.2cm}}
The regulating factor $R(\ell)$ suppresses all IR-divergent regions (for all integrands), making the result manifestly IR-finite. Ultimately, this works because all infrared singularities in a planar integral arise from predetermined integration regions---namely, the collinear regions \mbox{$\ell\!\to\!\alpha\,x_a\pl\,(1\mi\,\alpha)\,x_{a+1}$}, where the factor $R(\ell)$ becomes $\mathcal{O}(\epsilon)$.

It is trivial to see that the regulated expression (\ref{definition_of_regulator}) produces the correct result for all finite observables in the limit of $\epsilon\!\to\!0$: since the factor \mbox{$R(\ell)\!\sim\!1\pl\,\mathcal{O}(\epsilon)$} everywhere except in the isolated regions responsible for  infrared divergences (where it is $\mathcal{O}(\epsilon)$), it can be ignored for any {\it convergent} integral.

We should emphasize a distinction we are making between {\it convergent} integrals and so-called ``finite'' integrals. Tautologically, a ``convergent'' integral is one which can be evaluated {\it without} regularization; this requires that it have {\it no} collinearly-divergent regions. This notion of convergence precludes the combinations of divergent integrals which happen to be ``finite'' in some particular regularization scheme (but for which integration {\it does} require regularization). The convergence of an integral can be tested as follows: multiply the integrand by any two adjacent propagators, and verify that the product vanishes in the corresponding collinear region:
\vspace{-0.2cm}\eq{\lim_{\delta\ell\to0}\left[\x{\ell}{a}\x{\ell}{b}\!\times\!\II(\ell\!=\!\alpha\, x_a+(1-\alpha)\,x_b+\delta\ell)\right]=0\quad\text{for }\quad b=a\pl1.\label{convergence_test}\vspace{-0.2cm}}
As pointed out in ref.\ \cite{ArkaniHamed:2010gh} the integrand for the ratio function is in fact {\it convergent}. A similar integrand-level test of (partial) convergence for the logarithm of the $4$-particle amplitude has been shown sufficient to completely fix the integrand through seven-loops, \cite{Bourjaily:2011hi,Eden:2011we}, and has also been used to find amplitudes involving more particles \cite{Golden:2012hi}.

It remains for us to show that the regulated amplitude (\ref{definition_of_regulator}) is actually given by (\ref{scalar_box_expansion_two_types_of_coefficients}). To see this, consider the box expansion as an integrand-level statement: we can decompose any one-loop {\it integrand} into parity-even and parity-odd sectors:
\vspace{-0.2cm}\eq{ \mathcal{A}_n^{(k),1}(\ell) = \sum_{\!\!\!1\leq a<b<c<d\!\!\!}\Big({\color{cut1}f^1_{a,b,c,d}}+{\color{cut2}f^2_{a,b,c,d}}\Big)\mathcal{I}_{a,b,c,d}(\ell)+ \text{parity-odd}.\label{scalar_boxes_and_parity_odd_expansion}\vspace{-0.2cm}}
(Recall that $\mathcal{I}_{a,b,c,d}(\ell)$ denotes the {\it integrand} of the scalar box $I_{a,b,c,d}$.) The scalar boxes form a complete basis for parity-even four-dimensional integrands (see e.g.\ \cite{vanNeerven:1983vr,Davydychev:1990cq,Bern:1993kr}), and so the first term of (\ref{scalar_boxes_and_parity_odd_expansion}) completely captures all parity-even contributions. The parity-odd contributions in (\ref{scalar_boxes_and_parity_odd_expansion}) are often ignored because they vanish when integrated over the parity-even contour of $\mathbb{R}^{3,1}$. Importantly, {\it all} parity-odd integrals are not merely vanishing upon integration, but are in fact {\it convergent} in the sense described above, \cite{ArkaniHamed:2010gh}. (This is not too surprising since the requirement for an integrand to vanish in the limit (\ref{convergence_test}) is itself {\it parity-invariant}.) And because all parity-odd integrands are convergent, the regulator $R(\ell)$ is $1\pl\,\mathcal{O}(\epsilon)$ everywhere, and can therefore be ignored.

For the parity-even sector---that is, the scalar box expansion, (\ref{scalar_box_expansion}) (but now understood at the {\it integrand}-level)---we only need to verify the following:
\vspace{-0.2cm}\eq{\int\!\! d^4\ell\,\, \frac{-\x{a}{c}\x{b}{d}\Delta\phantom{-}}{\x{\ell}{a}\x{\ell}{b}\x{\ell}{c}\x{\ell}{d}} R(\ell)=\int\!\! d^4\ell\,\, \frac{-\x{a}{c}\x{b}{d}\Delta\phantom{-}}{\x{\ell}{\hat{a}}\x{\ell}{\hat{b}}\x{\ell}{\hat{c}}\x{\ell}{\hat{d}}}+ \mathcal{O}(\epsilon)=I^{\epsilon}_{a,b,c,d}.\vspace{-0.2cm}}
This identity is proved by noting that all other factors in $R(\ell)$ are approximately unity except in very small regions, where the unregulated box is not singular for lack of divergent propagators; with the explicit propagators regulated, the singular regions are all removed, resulting in precisely $I^{\epsilon}_{a,b,c,d}$ as described in \mbox{section \ref{scalar_box_integrals_subsection}} and given in \mbox{Table \ref{dci_regulated_box_integrals}}. This concludes our proof that (\ref{scalar_box_expansion_two_types_of_coefficients}), using the `DCI'-regulated scalar box integrals, correctly reproduces all regulator-independent contributions to loop amplitudes, and therefore leads to correct formulae for all finite one-loop observables.

\subsection{Generalization of `DCI'-Regularization to Higher Loop-Orders}\label{higher_loop_regularization_section}

The failure of simple off-shell regularization beyond one-loop has been known for quite some time (see e.g.\ \cite{Drummond:2006rz}). For example, for the two-loop $4$-particle amplitude in $\mathcal{N}\!=\!4$ SYM, a simple off-shell prescription fails even to give the correct
coefficient for the double-logarithmic divergence! Because the `DCI'-regulator (\ref{definition_of_regulator}) is somewhat similar to an off-shell regulator at one-loop (but one involving non-uniform masses), this may seem like a bad omen for extending it beyond one-loop. However, the most natural generalization of (\ref{definition_of_regulator}) beyond one-loop {\it cannot} be interpreted as an off-shell regulator---which is good news, indeed! Let us first describe the generalization of the `DCI'-regulator to higher loop-orders, and then illustrate how it differs from an of-shell regulator in the case of $4$-particles.

The integrand-level understanding of the `DCI'-regulator, (\ref{definition_of_regulator}), provides an obvious generalization to higher loops-orders,
\vspace{-0.2cm}\eq{\int\limits_\textrm{reg}\!\!\prod_{i=1}^Ld^4\ell_i\,\, \II(\ell_1,\ldots,\ell_L) \equiv \int \prod_{i=1}^L\left(d^4\ell_i~R(\ell_i) \right) ~\II(\ell_1,\ldots,\ell_L)\,.\label{multiloop_regulator}\vspace{-0.2cm}}
We suspect that this regularization prescription will render all multi-loop integrands IR-finite; moreover, using the same arguments as in \mbox{section \ref{proof_of_dci_regularization_scheme_section}}, we expect that (\ref{multiloop_regulator}) will generate the correct result for all finite multi-loop observables.

As one simple example of a regularized integral beyond one-loop---and as an illustration of the differences between (\ref{multiloop_regulator}) and off-shell regularization---consider the `DCI'-regularized $4$-particle double-box integrand, \cite{Bern:1997nh}:
\eq{\begin{split}\hspace{-0.0cm}I^{\epsilon}_\textrm{reg}\!\equiv& \int\!\!d^4\ell_1d^4\ell_2\,\,\frac{\x{1}{3}\x{2}{4}}{\x{\ell_1}{1}\x{\ell_1}{2}\x{\ell_1}{3}\x{\ell_1}{\ell_2}\x{\ell_2}{3}\x{\ell_2}{4}\x{\ell_2}{1}} \,\, R(\ell_{1\phantom{2}}\!\!)R(\ell_{2\phantom{1}}\!\!)\,,\\=& \int\!\!d^4\ell_1d^4\ell_2\,\, \frac{\x{1}{3}\x{2}{4}}{\x{\ell_1}{\hat 1}\x{\ell_1}{\hat 2}\x{\ell_1}{\hat 3}\x{\ell_1}{\ell_2}\x{\ell_2}{\hat 3}\x{\ell_2}{\hat 4}\x{\ell_2}{\hat 1}}\,\, \frac{\x{\ell_1}{4}\x{\ell_2}{2}}{\x{\ell_1}{\hat 4}\x{\ell_2}{\hat 2}}\,.\end{split}\label{dci_regulated_double_box}}
Because of the part of $R(\ell_1)R(\ell_2)$ which survives, this is clearly {\it not} an off-shell version of the double-box!
In particular, the integrand (\ref{dci_regulated_double_box}) includes regions (both collinear and soft-collinear) where the regularization-factor {\it cannot} be approximated by unity.  This is good news, since this had to happen for the regulator to have any chance of working beyond one-loop.

It would be interesting to compute this integral explicitly, and verify for example that scheme-independent quantities such as the two-loop double-logarithmic divergences (the so called cusp anomalous dimension) are correctly reproduced; we leave this, however, to future work.

\subsection{The Infrared-Divergences of One-Loop Amplitudes}\label{cancelation_of_IR_divergences_section}
Let us now consider how the IR-divergences of scattering amplitudes are organized in the `DCI'-regulated box expansion of (\ref{scalar_box_expansion_two_types_of_coefficients}). These divergences arise from the parts of $I^{\epsilon}_{a,b,c,d}$ proportional to $\log(\epsilon)$ or $\log(\epsilon)^2$; let us denote the combined coefficients of each of these divergences as follows:
\vspace{-.2cm}\eq{\int\!\! d^4\ell\,\,\mathcal{A}_n^{(k),1}\equiv-F_2\log(\epsilon)^2-F_1\log(\epsilon)-F_0+\mathcal{O}(\epsilon).\vspace{-.2cm}\label{scalar_box_divergences}}

It is easy to identify the coefficient $F_2$ from \mbox{Table \ref{dci_regulated_box_integrals}}:\footnote{We should note that the coefficient of $\log(\epsilon)^2$ in the amplitude is \emph{twice} the coefficient of $\log(\mu_\textrm{IR}^2)^2$ of the Higgs regulator described in ref.\ \cite{Alday:2009zm}---equivalently, it is \emph{twice} the coefficient of $1/(2(D-4)^2)$ appearing in dimensional-regularization. This can be understood from the way (\ref{definition_of_regulator}) cuts-out the soft-collinear divergent regions of integration.} the only integrals which include a factor of $\log(\epsilon)^2$ are the so-called `two-mass hard' and `one-mass' boxes,
\vspace{-0.2cm}\eq{F_2=\sum_{a+2<d\leq n}\Big({\color{cut1}f^1_{a,a+1,a+2,d}}+{\color{cut2}f^2_{a,a+1,a+2,d}}\Big)=n\mathcal{A}_n^{(k),0}\vspace{-0.2cm}\label{scalar_box_expansion_integrated_F2}}
which can be understood\footnote{The factor of $n$ in (\ref{scalar_box_expansion_integrated_F2}) should be thought of as `$2n/2$', arising from the fact that each two-mass hard box diverges like $\frac{1}{2}\log(\epsilon)^2$ (one-mass boxes counted twice by symmetry), and the quad-cut coefficients ${\color{cut1}f^1}$ and ${\color{cut2}f^2}$ {\it separately} generate the tree-amplitude for each of the $n$ sets of adjacent legs.} as the `BCF'  representation of tree-amplitudes discovered in ref.\ \cite{Britto:2004ap} (see also \cite{Roiban:2004ix}).

Interestingly, the coefficient $F_1$ in (\ref{scalar_box_divergences}) is {\it also} proportional to the tree-amplitude. (This follows from the ideas discussed in \mbox{section \ref{IR_equations_section}}.) Because the coefficient of $\log(\epsilon)$ must be of transcendentality-one, it should involve the logarithm of some dual-conformal cross ratio, which we denote $\Omega_n$; ultimately, $F_1$ is found to be simply,\\[-4pt]
\vspace{-0.2cm}\eq{F_1=2\,\mathcal{A}_n^{(k),0}\log(\Omega_n)\quad\mathrm{with}\quad\Omega_n\equiv\prod_{a}\frac{\x{a}{a\pl2}}{\x{a}{a\pl3}}.\vspace{-.25cm}}

An important object that arises in the discussion of N${}^k$MHV amplitudes $\mathcal{A}_n^{(k),\ell}$ is the so-called {\it ratio function}, denoted $\mathcal{R}_n^{(k),\ell}$, which at one-loop is given by,
\vspace{-0.2cm}\eq{\mathcal{R}_n^{(k),1}\equiv\mathcal{A}_n^{(k),1}-\mathcal{A}_n^{(k),0}\!\times\!\mathcal{A}_n^{(0),1}.\vspace{-0.2cm}}
Because the divergences $F_1$ and $F_2$ are always proportional to the tree-amplitude, $\mathcal{A}_n^{(k),0}$---and because in momentum-twistor space $\mathcal{A}_n^{(0),0}$ is simply the identity---we see that $\mathcal{R}_n^{(k),1}$ is finite in the limit $\epsilon\!\to\!0$. Therefore, any one-loop ratio function can be computed more simply as,
\vspace{-0.0cm}\eq{\boxed{\int\!\!d^4\ell\,\,\mathcal{R}_n^{(k),1}=\sum_{\!\!\!1\leq a<b<c<d\!\!\!}\Big({\color{cut1}f^1_{a,b,c,d}}+{\color{cut2}f^2_{a,b,c,d}}-\mathcal{A}_n^{(k),0}\!\times\!f^{\text{(MHV)}}_{a,b,c,d}\Big)I^{\mathrm{fin}}_{a,b,c,d},}\vspace{-0.1cm}\label{finite_box_expansion_ratio_function}}
where $f^{\text{(MHV)}}_{a,b,c,d}$ are the MHV one-loop box coefficients, and $I^{\mathrm{fin}}_{a,b,c,d}$ are the $\epsilon$-independent (``finite'') parts of the $\epsilon$-regulated scalar box integrals $I^{\epsilon}_{a,b,c,d}$ as listed in \mbox{Table \ref{dci_regulated_box_integrals}}. Notice that because $I^{\epsilon}_{a,b,c,d}$ depended {\it only} on $\epsilon$ and DCI cross-ratios, the integrals $I^{\mathrm{fin}}_{a,b,c,d}$ are {\it manifestly} dual-conformally invariant! And so (\ref{finite_box_expansion_ratio_function}) provides a {\it manifestly} dual-conformal representation of any one-loop ratio function in $\mathcal{N}\!=\!4$ SYM.

\subsection{Theories with Triangle Contributions}\label{triangles_discussion}
We should briefly mention that the regulator (\ref{definition_of_regulator}) can be applied to any planar theory, not just $\mathcal{N}=4$ SYM. (Planarity being a consequence of the way the momenta $p_a$ are associated with region-momentum coordinates $x_a$; a possible generalization to the non-planar case will be discussed shortly.)

In a more general planar field theory, one-loop amplitudes may also require contributions from `triangle' and `bubble' integrals in addition to the scalar boxes. These integrals manifestly break dual-conformal symmetry, but at least the triangles can be regulated in the same way as the boxes, (\ref{definition_of_regulator}). Indeed, regulated triangle integrals can be obtained from the `DCI'-regulated scalar-box integrals without any additional work: one need only to send one of the points $x_a,\ldots,x_d$ to (space-like) `infinity', denoted $x_{\infty}$. The correctness of this is easily seen from the geometry of the four-mass integral, (\ref{def_four_mass_box}). Thus, for example, the (no-longer `DCI', but) $\epsilon$-regulated two-mass triangle integral can be found by simply sending a point of the three-mass integral to infinity:
\vspace{-0.2cm}\eq{\signA I^{\epsilon}_{a,a+1,c,d}=\signD\Li(1-v)\signB\frac{1}{2}\log(u')\log(v)\signB\frac{1}{2}\log(\epsilon)\log(v)+\mathcal{O}(\epsilon),\vspace{-0.2cm}}
where we simply take $x_d\!\to\!x_\infty$ so that,
\vspace{-0.2cm}\eq{\hspace{-2cm}u'\!\equiv\!\lim_{d\to\infty}\!\left\{\!\frac{\x{a\,\mi\,1}{b}\x{a}{b\pl1}\x{c}{d}}{\x{a\,\mi\,1}{b\pl1}\x{a}{c}\x{b}{d}}\!\right\}\!=\!\frac{\x{a\,\mi\,1}{b}\x{a}{b\pl1}}{\x{a\,\mi\,1}{b\pl1}\x{a}{c}},\quad v\!\equiv\!\lim_{d\to\infty}\!\left\{\!\frac{\x{b}{c}\x{a}{d}}{\x{a}{c}\x{b}{d}}\!\right\}\!=\! \frac{\x{b}{c}}{\x{a}{c}}\,.\hspace{-2cm}\nonumber\vspace{-0.2cm}}
Here, dual-conformal invariance is broken {\it explicitly} by the fact that `$d\!\to\!\infty$' picks out a preferred point---namely, $x_\infty$---in region-momentum space. The so-called `bubble' and rational terms are unaffected by the infrared regulator.

Our claim is that in any planar quantum field theory requiring triangle-contributions (which are absent for $\mathcal{N}\!=\!4$), these contributions are correctly reproduced by simply augmenting the box expansion with triangle integrals obtained from $I^{\epsilon}_{a,b,c,d}$ as described above, with coefficients fixed by on-shell diagrams. The inclusion of bubble and rational terms would be unchanged from their usual form (see e.g.\ \cite{Bern:2007dw}).

\subsection{Integrand-Level Infrared Equations and Residue Theorems}\label{IR_equations_section}

The well-understood factorization structure of infrared divergences of one-loop amplitudes in gauge theory \cite{Giele:1991vf,Kunszt:1994np} leads to constraints on integral coefficients in the framework of generalized unitarity, resulting in the so-called `infrared (IR)  equations' (see refs. \cite{Roiban:2004ix,Britto:2004ap,Bern:2004ky} for a few applications). These equations can be understood in terms of unitarity cuts, and in the planar case lead to $n(n\mi3)/2$ relations among the integral coefficients.

By considering generalized unitarity at the {\it integrand} level, however, the factorization structure of soft-collinear divergences leads to more powerful identities. Recall that the box coefficients are so-called `quad-cuts' (co-dimension four residues) of the loop integrand; let us now consider `triple-cuts'---co-dimension three residues. Because the loop integral is four-dimensional, a triple-cut still depends on one integration variable; by applying Cauchy's residue theorem to the integral over this remaining variable, we find that the sum of all box coefficients sharing a triple-cut must vanish\footnote{We should mention that there can be residues of poles at infinity---the parity-even combinations of such are simply the so-called triangle coefficients; the inclusion of these terms would have no substantive effect on the results of this section.}.

The richest of these residue theorems arise for triple-cuts involving at least one massless corner, as these separate into two distinct classes depending on whether the $3$-particle amplitude at the massless corner is $\mathcal{A}_3^{(-1)}$ or $\mathcal{A}_3^{(0)}$. Importantly, Cauchy's theorem applies to these two cases {\it separately}, leading to a {\it pair} of identities:\\[-6pt]
\vspace{-0.2cm}\eq{\begin{array}{lll} \displaystyle\sum_{e=d{+}1}^{a-1} {\color{cut1}f^1_{a,b,d,e}} -\sum_{c=b+1}^{d-1} {\color{cut1}f^1_{a,b,c,d}}&= \tau_{a,b,d}\,\mathcal{A}_n^{(k),0};\\
\displaystyle \sum_{e=d{+}1}^{a-1} {\color{cut2}f^2_{a,b,d,e}} -\sum_{c=b+1}^{d-1} {\color{cut2}f^2_{a,b,c,d}}&= \tau_{a,b,d}\,\mathcal{A}_n^{(k),0};\end{array}\quad \mathrm{(} b\!=\!a\pl1)\vspace{-0.1cm}\label{residue_theorems}}
where $\tau_{a,b,d}\!=\!\pm1$ if $d\!=\!b\pl\,1$ or $d\!=\!a\,\mi\,1$, respectively, and vanishes otherwise. The sums appearing in the left-hand side of (\ref{residue_theorems}) are over all box coefficients sharing a particular (chiral) triple-cut, while the right-hand sides follow from the universal structure of soft-collinear divergences (and when $\tau\!\neq\!0$, (\ref{residue_theorems}) simply represents the famous `BCF' formula for tree-amplitudes, \cite{Britto:2004ap,Britto:2004nc}).

It is easy to see that these constitute $2n(n\,\mi\,3)$ linearly-independent equations. Averaging the two lines of (\ref{residue_theorems}) results in $n(n\,\mi\,3)$ relations among the (parity-even) box coefficients---which is twice as many as the ``standard'' IR-equations. This doubling of equations is a consequence of the fact that each collinear divergence of the integrated amplitude can arise from {\it two} distinct integration regions; and requiring factorization at the {\it integrand} level results in identities for each integration region separately.

To better understand the nature of the vanishing terms appearing on the right-hand side of (\ref{residue_theorems}), let us consider a loop-integrand in the neighborhood of a soft-collinear divergence---where three consecutive propagators are simultaneously put on-shell. For the sake of concreteness, we may parameterize the loop integrand in such a region by writing $\ell\!\to\!(L^+,L^-,\ell_{\perp})$ (see \cite{ArkaniHamed:2008gz} for a similar discussion):\\[-4pt]
\vspace{-0.2cm}\eq{\II(\ell)\to \mathcal{A}_n^{(k),0}\int\!\! \frac{dL^-dL^+d^2\ell_\perp}{((L^+\!+1)L^-\!-\ell_\perp^2)(L^+L^-\!-\ell_\perp^2)(L^+(L^-\!+1)-\ell_\perp^2)}\,\II'(\ell)\,. \label{integrand_level_factorization_2}\vspace{-0.1cm}}
We can take a residue in $L^-$ which puts the first propagator on-shell, resulting in
\vspace{-0.2cm}\eq{\mathcal{A}_n^{(k),0}\int\!\! \frac{dL^+}{L^+-\ell_\perp^2} \frac{d^2\ell_\perp}{\ell_\perp^2}\,\II'(\ell)\label{integrand_level_factorization_2}\vspace{-0.2cm}}
(for the regime in which $L^+$, $\ell_\perp$ are both small). If we now write $\ell_\perp^2\!=\!z\tilde z$ and drop the condition that $\tilde z$ is the complex conjugate of $z$,
we see that we can take two more residues---for instance $L^+$ and then $z$---to localize to the triple-cut, and then finally take a residue involving the pole in $\tilde{z}$. Notice that this results in a `quad-cut' (a co-dimension four residue) which only involves {\it three} propagators! This is a simple one-loop example of the phenomenon of `composite leading singularities' discussed in refs.\ \cite{Cachazo:2008vp,Buchbinder:2005wp}, and is the physical origin of the terms on the right-hand side of (\ref{residue_theorems}).

It is not hard to show that the criterion for the convergence of an integral discussed in \mbox{section \ref{proof_of_dci_regularization_scheme_section}} is equivalent to the requirement that {\it all} composite leading singularities vanish (in general, composite leading singularities are in one to one correspondence with IR-divergent triangle topologies, which are in correspondence with these equations); this implies that for ratio functions in $\mathcal{N}\!=\!4$ SYM the right-hand sides of (\ref{residue_theorems}) must vanish. Moreover, it is possible to show that the difference between the `DCI'-regularized boxes of (\ref{finite_box_expansion_ratio_function}), and more familiar---e.g.\ dimensionally-regulated--- expressions for the box integrals is necessarily proportional to the the right-hand-sides of (\ref{residue_theorems})---which provides an alternate proof of the equivalence between the two regularization schema when applied to manifestly finite observables. We leave the details of this discussion as an exercise for the interested reader.

In summary, the residue theorems (\ref{residue_theorems}) encode the physical fact that IR-factorization occurs at the {\it integrand} level, and constitute $n(n\,\mi\,3)$ independent parity-even relations among the box coefficients---{\it twice} as many as the IR-equations arising after integration. These relations are {\it general}---that is, not specific to $\mathcal{N}\!=\!4$ SYM---provided that triangle coefficients are included where they are required.

\subsection{Applications to Non-Planar Theories}\label{non_planar_generalization_section}

Although we have largely focused our attention on planar theories, it is worth emphasizing that the increased number of IR equations upon considering integrands instead of integrals is completely unrelated to planarity:  the same result holds for non-planar theories as well---including gravity. This motivates a generalization of the integrand-level regulator (\ref{definition_of_regulator}) to the non-planar case, at least at one-loop order.

As a simple illustration, consider the $4$-particle amplitude in $\mathcal{N}\!=\!8$ supergravity.  By the no-triangle property, \cite{BjerrumBohr:2008ji}, it is a combination of at most three box functions. It is not hard to see that the absence of collinear divergences---the vanishing of the right-hand-sides of the analogs of (\ref{residue_theorems})---uniquely fixes their relative coefficients, leading to the following form for the amplitude:\\[-4pt]
\vspace{-0.1cm}\eq{\mathcal{A}_4^{(0),1;\mathcal{N}=8}\propto u F(p_1,p_2,p_3,p_4)+s F(p_1,p_3,p_2,p_4)+t F(p_1,p_3,p_4,p_2),\vspace{-0.0cm}}
where $s,t,u$ are the usual Mandelstam invariants and the box integrals are labelled by the momenta coming into the vertices. It is not difficult to check that this matches the correct expression. As a less trivial example, consider the 6-graviton amplitude (with arbitrary heliciticies). In principle, this amplitude could involve any combination of 195 boxes; but the integrand-level IR-equations of the preceding subsection shows that the amplitude can be expressed in terms of at most 120 combinations. That is, we find 75 linearly-independent constraints of the form (\ref{residue_theorems}), as opposed to the mere 25 constraints arising from the previously known IR-equations. It would be interesting to explore if these relations could be used to obtain more compact analytic forms for one-loop graviton amplitudes.

\section{The {\it Chiral} Box Expansion for One-Loop {\it Integrands}}\label{chiral_box_expansion_section}
As we have seen, while the familiar box expansion reproduces all one-loop amplitudes {\it post-integration}, it does {\it not} match the full structure of the actual loop integrand. These can be easily understood in the case of MHV loop amplitudes, where the {\it actual} MHV loop integrand {\it only} has support on two-mass easy boxes involving ${\color{cut1}\ell_1}$---with vanishing support on all quad-cuts involving ${\color{cut2}\ell_2}$. However, because the scalar box integrals are {\it parity-even}, their integrands always have unit-magnitude residues for both quad-cuts. This is easy to understand, as scattering amplitude integrands are generally {\it chiral}, while the scalar boxes are manifestly {\it non-chiral}.

In this section, we describe a slight modification to the scalar-box integrands given above which leads to a fully-chiral generalization of the box expansion, allowing us to represent all one-loop {\it integrands} of $\mathcal{N}\!=\!4$ SYM. In fact, such a modification was discovered for MHV and NMHV one-loop integrands in ref.\ \cite{ArkaniHamed:2010gh}, but the generalization to more complicated amplitudes was unclear. Here, by revisiting the special case of MHV, we will find that the underlying structure naturally generalizes to {\it all} N$^{k}$MHV one-loop integrands, $\mathcal{A}_n^{(k),1}$.

All of the essential structure needed is already present in the case of the MHV one-loop integrands, $\mathcal{A}^{(0),1}_n$. In ref.\ \cite{ArkaniHamed:2010gh}, the one-loop MHV integrand was written:
\vspace{-.2cm}\eq{\hspace{-2.cm}\begin{array}{c}
\displaystyle\text{{\Large$\displaystyle\sum_{a<c<a}$}}\raisebox{-38pt}{\includegraphics[scale=1]{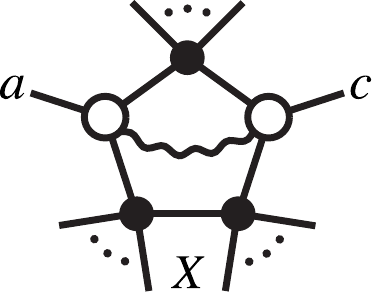}}\!\!\!\equiv\text{{\Large$\displaystyle\sum_{a<c<a}$}}\frac{\ab{\ell\,(a\mi1\,a\,a\pl1)\!\newcap\!(c\mi1\,c\,c\pl1)}\ab{X\,c\,a}}{\ab{\ell\,a\mi1a}\ab{\ell\,aa\pl1}\ab{\ell\,c\mi1c}\ab{\ell\,cc\pl1}\ab{\ell\,X}}.\\[-0pt]\end{array}\hspace{-1.5cm}\vspace{-.2cm}\label{pentagon_expansion_for_mhv_one_loop}}
Here, $X$ is an arbitrary, auxiliary line in momentum-twistor space (of course, the integrand is ultimately, algebraically independent of $X$). We will mostly take this formula for granted here; but let us see what role is played by the auxiliary line $X$, and how we may generalize this to reproduce any one-loop integrand.

Because a pentagon integral has five propagators, it has $2\!\times\!\binom{5}{4}\!=\!10$ fourth-degree residues---from the two ways of cutting any four of the five propagators. But because one of its propagators involves this auxiliary line $X$, only two such residues are physically meaningful: those which cut the lines $\{(a\mi1\,a),(a\,a\pl1),(c\mi1\,c),(c\,c\pl1)\}$. These are obviously `two-mass easy' quad-cuts; but because the numerator of the integrand in (\ref{pentagon_expansion_for_mhv_one_loop}) is proportional to $\ab{\ell\,(a\mi1\,a\,a\pl1)\!\newcap\!(c\mi1\,c\,c\pl1)}$, the pentagon's residue on ${\color{cut2}\ell_2}$ vanishes, while the residue on ${\color{cut1}\ell_1}$ is unity (see \mbox{Table \ref{one_loop_leading_singularities_table}}). But this is perfect: the one-loop MHV integrand only has support on ${\color{cut1}\ell_1}$!

Because of this, we choose to view the {\it pentagon} contributions to (\ref{pentagon_expansion_for_mhv_one_loop}) as something like a `chiralized' version of the scalar two-mass easy box:
\vspace{-.2cm}\eq{{\color{cut1}\tilde{\mathcal{I}}^1_{a,a+1,c,c+1}}\equiv\raisebox{-48pt}{\includegraphics[scale=1]{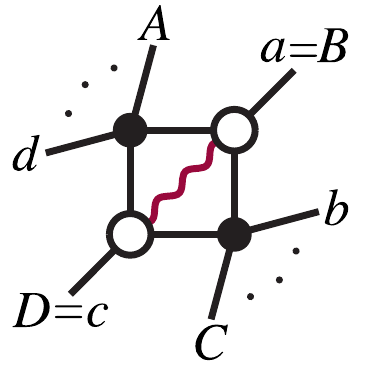}}\equiv \frac{\ab{\ell\,(a\mi1\,a\,a\pl1)\!\newcap\!(c\mi1\,c\,c\pl1)}\ab{X\,c\,a}}{\ab{\ell\,a\mi1a}\ab{\ell\,aa\pl1}\ab{\ell\,c\mi1c}\ab{\ell\,cc\pl1}\ab{\ell\,X}}.\vspace{-.2cm}\label{chiral_two_mass_easy_one}}
We are motivated to draw this as a {\it box} because it has precisely {\it one} physically-meaningful quad-cut (in this case ${\color{cut1}\ell_1}$), upon which it has residue $\mi1$. Although not relevant for MHV integrands, we could similarly define the parity-conjugate version,
\vspace{-.2cm}\eq{{\color{cut2}\tilde{\mathcal{I}}^2_{a,a+1,c,c+1}}\equiv\raisebox{-48pt}{\includegraphics[scale=1]{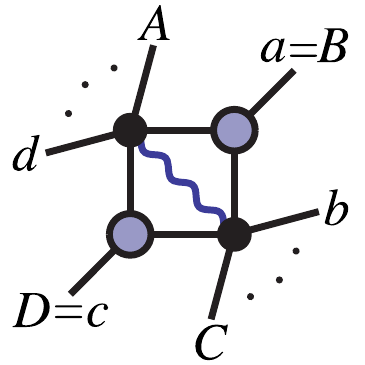}}\equiv \frac{\ab{\ell\,c\,a}\ab{X\,(a\mi1\,a\,a\pl1)\!\newcap\!(c\mi1\,c\,c\pl1)}}{\ab{\ell\,a\mi1a}\ab{\ell\,aa\pl1}\ab{\ell\,c\mi1c}\ab{\ell\,cc\pl1}\ab{\ell\,X}},\vspace{-.2cm}}
which has residue of $\mi1$ on the quad-cut ${\color{cut2}\ell_2}$, and vanishing residue on ${\color{cut1}\ell_1}$.

Although it may seem like we're nearly done, we must step back to observe that {\it not all the terms in (\ref{pentagon_expansion_for_mhv_one_loop}) are pentagons!} This is indeed a good thing, because as described in ref.\ \cite{ArkaniHamed:2010gh}, all such chiral pentagons are {\it convergent}, while the actual MHV one-loop amplitude is of course divergent! The easy-to-overlook, non-pentagon contributions to the one-loop MHV integrand of (\ref{pentagon_expansion_for_mhv_one_loop}), come from the boundary terms when $c\!=\!a\pl1$:
\vspace{-.2cm}\eq{\hspace{-3cm}\begin{array}{lll} \mathcal{I}^{\mathrm{div}}_a\equiv\hspace{-10pt}\raisebox{-38pt}{\includegraphics[scale=1]{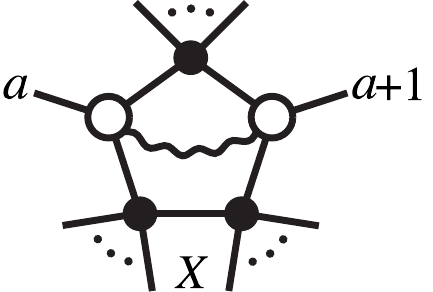}}\hspace{-22pt}&\displaystyle=\frac{\ab{\ell\,(a\mi1\,a\,a\pl1)\!\newcap\!(a\,a\pl1\,a\pl2)}\ab{X\,a\pl1\,a}}{\ab{\ell\,a\mi1a}{\color{hteal}\ab{\ell\,aa\pl1}\ab{\ell\,aa\pl1}}\ab{\ell\,a\pl1a\pl2}\ab{\ell\,X}},\\[-22pt]&=\displaystyle\frac{{\color{hteal}\ab{\ell\,aa\pl1}}\ab{a\mi1aa\pl1a\pl2}\ab{X\,a\pl1\,a}}{\ab{\ell\,a\mi1a}{\color{hteal}\ab{\ell\,aa\pl1}\ab{\ell\,aa\pl1}}\ab{\ell\,a\pl1a\pl2}\ab{\ell\,X}},\\[-20pt]&\displaystyle=\frac{\ab{a\mi1aa\pl1a\pl2}\ab{X\,a\pl1\,a}}{\ab{\ell\,a\mi1a}\ab{\ell\,aa\pl1}\ab{\ell\,a\pl1a\pl2}\ab{\ell\,X}}&\hspace{-20pt}\equiv\hspace{-15pt}\raisebox{-38pt}{\includegraphics[scale=1]{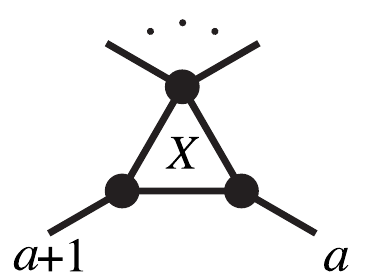}}\\[-55pt]\end{array}\hspace{-3cm}\vspace{45pt}}
We are motivated to draw this as a triangle, because it has only three non-$X$ propagators. In fact, in the particular case where $X$ is taken to be the point at infinity, $\mathcal{I}^{\mathrm{div}}_a$ becomes precisely a scalar triangle.

Therefore, although somewhat less concise than
(\ref{pentagon_expansion_for_mhv_one_loop}) (which is deceptively so), we can write the MHV one-loop {\it integrand} in the somewhat more suggestive form,
\vspace{-.2cm}\eq{\mathcal{A}_n^{(0),1}=\sum_{\!\!\!1\leq a<c\leq n\!\!\!}\Big({\color{cut1}\tilde{\mathcal{I}}^1_{a,a+1,c,c+1}}\Big)+\sum_a\mathcal{I}^{\mathrm{div}}_a\,,\vspace{-.2cm}\label{chiral_box_expansion_for_mhv}}
where the divergent part is expressed solely in terms of triangles.
(Although we have not yet defined {\it all} chiral boxes ${\color{cut1}\tilde{\mathcal{I}}^1_{a,b,c,d}}$, the two-mass easy boxes given in (\ref{chiral_two_mass_easy_one}) are the only ones relevant for MHV ($k\!=\!0$).)

From this simple example, it should be clear that if we have `chiral' versions of all the scalar boxes---ones with precisely {\it one} physical quad-cut with residue $\mi1$ on either ${\color{cut1}\ell_1}$ and ${\color{cut2}\ell_2}$---then would have a `chiral' version of the box expansion:
\vspace{-.2cm}\eq{\boxed{\mathcal{A}_n^{(k),1}=\sum_{\!\!\!1\leq a<b<c<d\!\!\!}\Big({\color{cut1}f^1_{a,b,c,d}\tilde{\mathcal{I}}^1_{a,b,c,d}}+{\color{cut2}f^2_{a,b,c,d}\tilde{\mathcal{I}}^2_{a,b,c,d}}\Big)+\mathcal{A}_n^{(k),0}\sum_a\mathcal{I}^{\mathrm{div}}_a.}\vspace{-.2cm}\label{chiral_box_expansion}}
Notice that this expression will be valid {\it before} integration---unlike the more familiar {\it scalar} box expansion described in \mbox{section \ref{generalized_unitarity_section}}. To see that this formula must be right, first observe that the chiral boxes with coefficients are specifically engineered to have {\it precisely} the same residues on all physical quad-cuts as the actual amplitude's integrand. However, as mentioned above, {\it every} chiral box integral $\tilde{I}^{{\color{cut1}1},{\color{cut2}2}}_{a,b,c,d}\!\equiv\!\int\!\!d^4\ell\,\,\widetilde{\mathcal{I}}^{{\color{cut1}1},{\color{cut2}2}}_{a,b,c,d}$ is {\it convergent}---and so the chiral boxes alone cannot fully represent the amplitude.
The remaining contributions must therefore encode the divergence of the one-loop amplitude, which from
(\ref{chiral_box_expansion_for_mhv}) together with integrand-level factorization is simply the sum of the divergent ``triangles'' $\mathcal{I}_{a}^{\mathrm{div}}$.

Thus, by construction, (\ref{chiral_box_expansion}) will have the correct leading singularities on all physical quad-cuts (those that do not involve $X$) and have the correct infrared divergences.   In order for (\ref{chiral_box_expansion}) to reproduce the {\it integrand}, it obviously must be $X$-independent; and so it must have vanishing support on all quad-cuts involving $X$.  The preceding arguments only fix the infrared-divergent part but not any potential four-mass integrals involving $X$. Thus, in order to uniquely fix ${\color{cut1}\tilde{\mathcal{I}}^1_{a,b,c,d}}$ and ${\color{cut2}\tilde{\mathcal{I}}^2_{a,b,c,d}}$, we must also require that they are {\it parity-odd} on all {\it four-mass} quad-cuts involving the auxiliary line $X$---that is, they must have {\it equal} residues (in both sign {\it and} magnitude) on all `spurious' quad-cuts (as any parity-even integrand must have {\it opposite} residues on parity-conjugate quad-cuts, this will ensure that no four-mass integrals involving $X$ will survive integration over a parity-invariant contour).

Given chiral box integrands ${\color{cut1}\tilde{\mathcal{I}}^1_{a,b,c,d}}$ and ${\color{cut2}\tilde{\mathcal{I}}^2_{a,b,c,d}}$ which satisfy the conditions described above---being infrared convergent, having only one nonvanishing physical quad-cut and being parity-odd on spurious quad-cuts--- it is not hard to prove that (\ref{chiral_box_expansion}) matches {\it all} physical residues of the actual loop integrand, that it is ultimately free of any quad-cuts involving $X$, and that it is moreover algebraically independent of $X$. And so, (\ref{chiral_box_expansion}) must give the correct integrand for the amplitude. (Using the {\sc Mathematica} package documented in \mbox{Appendix \ref{mathematica_implementation_section}}, it is easy to verify that (\ref{chiral_box_expansion}) directly matches the one-loop integrands obtained using the BCFW recursion relations (which is described in \mbox{Appendix \ref{bcfw_loop_recursion_section}}).)

\begin{table}[t*]\caption{{\bf Chiral One-Loop `Box' Integrands and Integrals}}\label{chiral_box_integrands_and_integrals_table}\vspace{-.3cm}
\noindent\scalebox{.815}{\mbox{\hspace{-0.0cm}\begin{minipage}[h]{\textwidth}\vspace{0cm}\eq{\hspace{-1.275cm}
\begin{array}{@{}c@{}}\begin{array}{|@{}c@{}|@{}c@{}|@{}c@{}|}\hline
\raisebox{-50.25pt}{\includegraphics[scale=1]{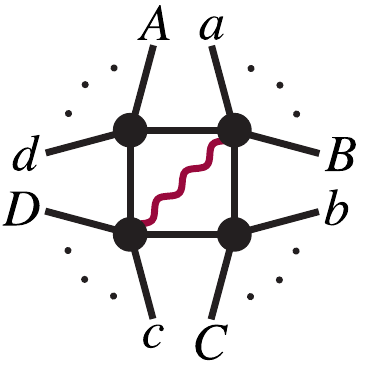}}&\begin{array}{@{}c||c@{}}\rule{191pt}{0.0pt}&\rule{191pt}{0.0pt}\\[-12pt]-\frac{1}{2}\ab{Aa\,Cc}\ab{Bb\,Dd}\ab{\ell\,(X)}\Delta&-\frac{1}{2}\ab{Aa\,Cc}\ab{Bb\,Dd}\ab{\ell\,(X)}\Delta\\
\mi\frac{1}{12}\epsilon^{ijkl}\ab{\ell\,(L_i)\!\newcap\!(L_jX_1)(L_k)\newcap(L_lX_2)}\!&\pl\frac{1}{12}\epsilon^{ijkl}\ab{\ell\,(L_i)\!\newcap\!(L_jX_1)(L_k)\newcap(L_lX_2)}\\[-10pt]
\\\hline\multicolumn{2}{c}{}\\[-10pt] \multicolumn{2}{@{}c@{}}{\mathrm{}\;\{L_1,L_2,L_3,L_4\}\equiv\{(Aa),(Bb),(Cc),(Dd)\},\quad \Delta\equiv\sqrt{(1\mi\,u\mi\,v)^2\mi\,4uv}}\\
\multicolumn{2}{@{}c@{}}{u\equiv u[a\,b;\!c\,d],\; v\equiv u[b\,c;\!d\,a],\quad\alpha\equiv\frac{1}{2}(1\mi\,u\,\pl\, v\pl\Delta),\; \beta\equiv\frac{1}{2}(1\pl\,u\,\mi\, v\pl\Delta)}\\\multicolumn{2}{c}{}\\[-10pt]
\hline
\multicolumn{2}{l}{\text{{\large$\begin{array}{l@{}l}\\[-37pt]\\\!\,\tilde{I}^{{\color{cut1}1},{\color{cut2}2}}\!=&-\Li(\alpha)\!-\!\Li(\beta)\!+\!\Li(1)\!-\!\frac{1}{2}\log(u)\log(v)\!+\!\log(\alpha)\log(\beta)\\[2pt]\end{array}$}}}\end{array}
&\raisebox{-50.25pt}{\includegraphics[scale=1]{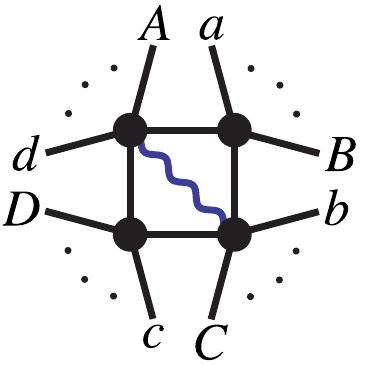}}
\\\hline\hline
\raisebox{-50.25pt}{\includegraphics[scale=1]{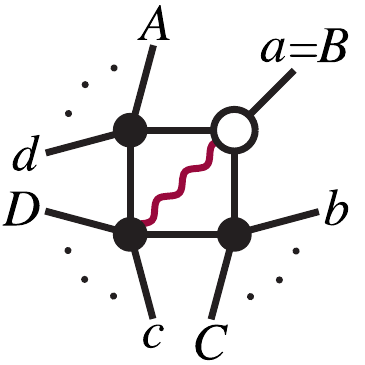}}&\begin{array}{@{}c||c@{}}\rule{191pt}{0.0pt}&\rule{191pt}{0.0pt}\\[-14pt]\!\!\!\frac{1}{2}\Big(\ab{\ell\,(Aab)\!\newcap\!(Cc(DdB)\!\newcap\!(X))}\phantom{\Big)}&\!\!\!\frac{1}{2}\Big(\ab{\ell\,B(Dd)\!\newcap\!((Cc)\!\newcap\!(Aab)(X))}\phantom{\Big)}\\
\!\!\!\phantom{\frac{1}{2}\Big(}\,\,\,\,\,\mi\ab{\ell\,(Aab)\!\newcap\!(Dd(CcB)\!\newcap\!(X))}\Big)&\!\!\!\phantom{\frac{1}{2}\Big(}\,\,\,\,\,\mi\ab{\ell\,B(Cc)\!\newcap\!((Dd)\!\newcap\!(Aab)(X))}\Big)\\[-14pt]
\\\hline\multicolumn{2}{l}{\text{{\large$\begin{array}{l@{}l}\\[-37pt]\\\!\,\tilde{I}^{{\color{cut1}1},{\color{cut2}2}}=&\phantom{+}\Li(1-u[c\,b;\!a\,X])-\Li(1-u[b\,c;\!d\,a])\\&+\Li(1-u[d\,a;\!b\,X])-\frac{1}{2}\log(u[c\,d;\!a\,X])\log(u[d\,a;\!b\,X])\\&-\frac{1}{2}\log(u[c\,b;\!a\,X])\log(u[d\,c;\!b\,X])\\[2pt]\end{array}$}}}\end{array}
&\raisebox{-50.25pt}{\includegraphics[scale=1]{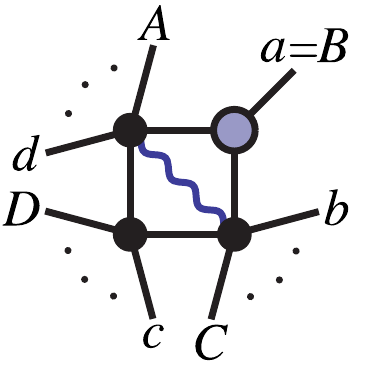}}
\\\hline\hline
\raisebox{-50.25pt}{\includegraphics[scale=1]{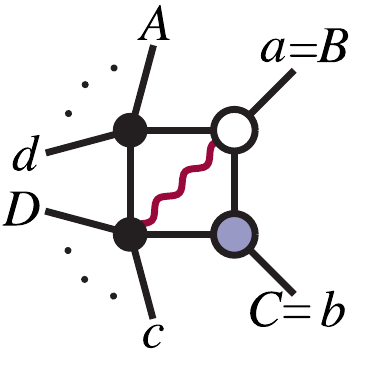}}&\begin{array}{@{}c||c@{}}\rule{191pt}{0.0pt}&\rule{191pt}{0.0pt}\\[-14pt]\!\!\!\frac{1}{2}\Big(\ab{\ell\,b(Aa)\!\newcap\!(Cc(DdB)\!\newcap\!(X))}\phantom{\Big)}&\!\!\!\frac{1}{2}\Big(\ab{\ell\,B(Cc)\!\newcap\!(Aa(Ddb)\!\newcap\!(X))}\phantom{\Big)}\\
\!\!\!\phantom{\frac{1}{2}\Big(}\,\,\,\,\,\mi\ab{\ell\,b(Aa)\!\newcap\!(Dd(CcB)\!\newcap\!(X))}\Big)&\!\!\!\phantom{\frac{1}{2}\Big(}\,\,\,\,\,\mi\ab{\ell\,B(Cc)\!\newcap\!(Dd(Aab)\!\newcap\!(X))}\Big)\\[-14pt]
\\\hline\multicolumn{2}{l}{\text{{\large$\begin{array}{l@{}l}\\[-37pt]\\\!\,\tilde{I}^{{\color{cut1}1},{\color{cut2}2}}=&\phantom{+}\Li(1-u[d\,c;\!b\,X])-\Li(1-u[d\,b;\!a\,X])
\\&+\frac{1}{2}\log(u[a\,b;\!c\,X])\log(u[d\,b;\!a\,X])\\&+\frac{1}{2}\log(u[a\,c;\!d\,X])\log(u[d\,c;\!b\,X])\\[2pt]\end{array}$}}}\end{array}
&\raisebox{-50.25pt}{\includegraphics[scale=1]{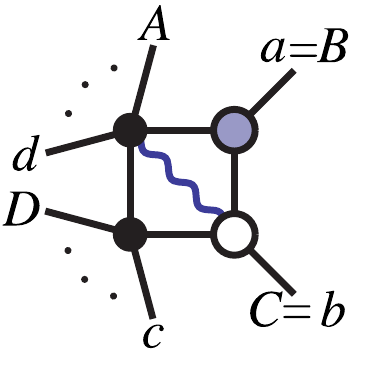}}
\\\hline\hline
\raisebox{-50.25pt}{\includegraphics[scale=1]{chiral_two_mass_easy_1}}&\begin{array}{@{}c||c@{}}\rule{191pt}{0.0pt}&\rule{191pt}{0.0pt}\\\ab{\ell\,(Aab)\!\newcap\!(Ccd)}\ab{XDB}&\ab{\ell\,DB}\ab{X(Aab)\!\newcap\!(Ccd)}\\
\\\hline\multicolumn{2}{l}{\text{{\large$\begin{array}{l@{}l}\\[-37pt]\\\!\,\tilde{I}^{{\color{cut1}1},{\color{cut2}2}}=&\phantom{+}\Li(1-u[d\,b;\!a\,X])-\Li(1-u[b\,c;\!d\,a]\,)\\&+\Li(1-u[a\,c;\!d\,X])-\Li(1-u[c\,b;\!a\,X])\\&+\Li(1-u[b\,c;\!d\,X])\,-\log(u[d\,b;\!a\,X])\log(u[a\,c;\!d\,X])\\[2pt]\end{array}$}}}\end{array}
&\raisebox{-50.25pt}{\includegraphics[scale=1]{chiral_two_mass_easy_2}}
\\\hline\hline
\raisebox{-50.25pt}{\includegraphics[scale=1]{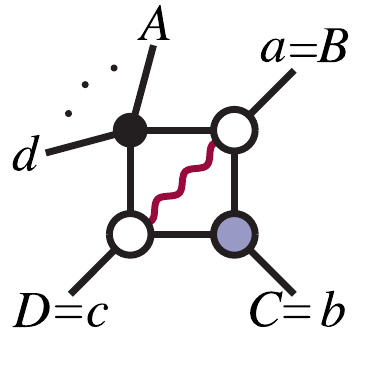}}&\begin{array}{@{}c||c@{}}\rule{191pt}{0.0pt}&\rule{191pt}{0.0pt}\\\ab{\ell\,(Aab)\!\newcap\!(Ccd)}\ab{XDB}&\ab{\ell\,DB}\ab{X(Aab)\!\newcap\!(Ccd)}\\
\\\hline\multicolumn{2}{l}{\text{{\large$\begin{array}{l@{}l}\\[-30pt]\\\!\,\tilde{I}^{{\color{cut1}1},{\color{cut2}2}}=&\phantom{+}\Li(1)-\Li(1-u[d\,b;\!a\,X])-\Li(1-u[a\,c;\!d\,X])\\&-\log(u[d\,b;\!a\,X])\log(u[a\,c;\!d\,X])\\&\\[-5pt]\end{array}$}}}\end{array}
&\raisebox{-50.25pt}{\includegraphics[scale=1]{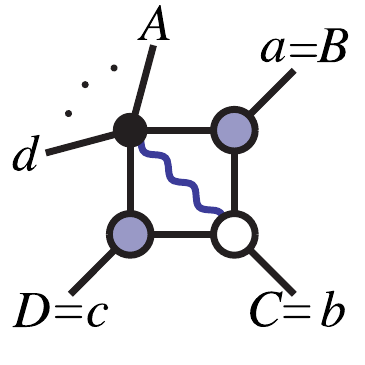}}
\\\hline\hline
\raisebox{-40.25pt}{\includegraphics[scale=1]{spurious_triangle}}&\begin{array}{@{}c||c@{}}&\\[-15pt]\rule{191pt}{0.0pt}&\rule{191pt}{0.0pt}\\[-10pt]\ab{a\,\mi\,1\,a\,\,a\pl1\,a\pl2}\ab{X\,\,a\pl1\,a}&\ab{a\,\mi\,1\,a\,\,a\pl1\,a\pl2}\ab{X\,\,a\pl1\,a}
\\[5pt]\hline\multicolumn{2}{l}{\mbox{\text{{\hspace{-0cm}\large$\begin{array}{l@{}l}\\[-15pt]\hspace{-0cm}\!\,\tilde{I}^{{\color{cut1}1},{\color{cut2}2}}\!=&-\Li(1)\!-\!\frac{1}{2}\log(u[a\mi1\,a\pl1;\!a\pl2\,X])\log(u[a\pl3\,a\pl1;\!a\,X])\hspace{-1cm}\\&-\frac{1}{2}\log(\epsilon)\log(u[a\mi1\,a\pl1;\!a\pl2\,X]u[a\pl3\,a\pl1;\!a\,X])\!-\!\frac{1}{2}\log(\epsilon)^2\hspace{-2cm}\end{array}$}\hspace{-1cm}}}}\end{array}
&\raisebox{-40.25pt}{\includegraphics[scale=1]{spurious_triangle}}
\\\hline\multicolumn{3}{c}{}\\[-8pt]
\multicolumn{3}{c}{\begin{array}{c}\displaystyle\mathrm{with}\qquad u[a\,b;\!c\,d]\equiv\frac{\ab{a\,\mi\,1\,a\,\,b\,\mi\,1\,b}\ab{c\,\mi\,1\,c\,\,d\,\mi\,1\,d}}{\ab{a\,\mi\,1\,a\,\,c\,\mi\,1\,c}\ab{b\,\mi\,1\,b\,\,d\,\mi\,1\,d}}\quad\mathrm{and}\quad u[a\,b;\!c\,X]\equiv\frac{\ab{a\,\mi\,1\,a\,\,b\,\mi\,1\,b}\ab{c\,\mi\,1\,c\,\,(X)}}{\ab{a\,\mi\,1\,a\,\,c\,\mi\,1\,c}\ab{b\,\mi\,1\,b\,\,(X)}}.\\[15pt]\displaystyle \phantom{\text{(The chiral box integrands satisfy }{\color{cut1}\widetilde{\mathcal{I}}^1_{a,b,c,d}}={\color{cut2}\widetilde{\mathcal{I}}^2_{b,c,d,a}}.)\qquad\qquad~}\\[-20pt]\end{array}}
\end{array}\end{array}\vspace{-.2cm}\nonumber\vspace{-1.2cm}}\end{minipage}}}\end{table}

In terms of the chiral boxes, the ratio function becomes,
\vspace{-.2cm}\eq{\boxed{\mathcal{R}_n^{(k),1}=\sum_{\!\!\!1\leq a<b<c<d\!\!\!}\Big({\color{cut1}f^1_{a,b,c,d}\tilde{\mathcal{I}}^1_{a,b,c,d}}+{\color{cut2}f^2_{a,b,c,d}\tilde{\mathcal{I}}^2_{a,b,c,d}}\Big)
-\mathcal{A}_n^{(k),0}\!\times\!\mathcal{A}_n^{(0),1}.}\vspace{-.2cm}\label{chiral_box_expansion_ratio}}
Notice that the divergent integrands $\mathcal{I}_a^{\mathrm{div}}$ are {\it manifestly} canceled in the ratio function, leading to an expression involving only {\it manifestly convergent} chiral boxes. This is remarkable, as it provides an analytic form of any one-loop ratio function for which no regularization is needed! However, although the complete integrand is of course independent of $X$, each chiral box {\it individually} depends on $X$. Nevertheless, although it is generally difficult (algebraically) to prove the $X$-independence of the integrated expressions (this can be seen to amount to the integrand-level IR equations (\ref{residue_theorems})), this remarkable fact can easily be verified numerically---for example, using the {\sc Mathematica} package `{\tt loop\uscore amplitudes}' which is made available with this note on the {\tt arXiv}, and documented in \mbox{Appendix \ref{mathematica_implementation_section}}.

~\newpage~\newpage
\section{Conclusions}~\\[-25pt]
In this note, we have revisited the familiar story of using generalized unitarity to reconstruct one-loop amplitudes, especially for the case of planar, $\mathcal{N}\!=\!4$ SYM. In order to make manifest all the known symmetries of the theory, we reconsidered the regularization of IR-divergences, and found a new, `DCI'-regularization scheme (\ref{definition_of_regulator}) which makes {\it manifest}---term-by-term---the dual-conformal invariance of all finite observables of $\mathcal{N}\!=\!4$ SYM at one-loop, including all N$^{k}$MHV ratio functions, (\ref{finite_box_expansion_ratio_function}).

The existence of such a regularization scheme was motivated by considering the remarkable properties of one-loop amplitudes {\it prior-to} integration. Such considerations also led to integrand-level IR-equations, (\ref{residue_theorems})---giving novel constraints among box coefficients, and having applications beyond the planar limit. And in \mbox{section \ref{chiral_box_expansion_section}}, we found that the familiar box expansion could be upgraded to a {\it chiral} box expansion for one-loop {\it integrands}, reproducing both the parity-odd and parity-even contributions to scattering amplitudes, and making the factorization of IR-divergences manifest.

For the sake of completeness and reference, we have comprehensively described all the ingredients required to compute any one-loop amplitude in planar $\mathcal{N}\!=\!4$ SYM: we gave explicit formulae for all `DCI'-regularized scalar box integrals in \mbox{Table \ref{dci_regulated_box_integrals}}, and we gave expressions for all one-loop box coefficients in \mbox{Table \ref{one_loop_leading_singularities_table}}. Remarkably, these tables are incredibly redundant: all degenerate cases of {\it both} tables follow smoothly from the generic case---from the four-mass integral, (\ref{symmetric_four_mass_integral}), and the four-mass functions, (\ref{four_mass_coefficient_1}) and (\ref{four_mass_coefficient_2}), respectively. 

The notation used throughout this paper is reviewed in \mbox{Appendix \ref{momentum_twistor_notation_and_review}}. In \mbox{Appendix \ref{bcfw_loop_recursion_section}}, we use the BCFW recursion relations described in \cite{ArkaniHamed:2010kv} to explicitly represent all one-loop integrands in $\mathcal{N}\!=\!4$ SYM. All the results described in this paper have been implemented in a {\sc Mathematica} package, `{\tt loop\uscore amplitudes}'; instructions for obtaining this package, and complete documentation of the functions made available by it are described in \mbox{Appendix \ref{mathematica_implementation_section}}.

Natural extensions of this work include applying the `DCI'-regulator to scalar integrals beyond one-loop, finding chiral representations of higher-loop integrands, and using the integrand-level IR-equations to find better representations of one-loop amplitudes beyond the planar limit.\\[-28pt]

\section*{Acknowledgements}~\\[-25pt]
We are especially grateful to Nima Arkani-Hamed and Freddy Cachazo for their helpful comments and suggestions regarding this work. This work was supported in part by the Harvard Society of Fellows and a grant form the Harvard Milton Fund (JB), Department of Energy contract \mbox{DE-FG02-91ER40654} (JB), and NSF grants \mbox{PHY-0756966} (JT) and \mbox{PHY-0969448} (SCH).

\newpage

\appendix
\section{Review of Momentum-Twistor Variables and Notation}\label{momentum_twistor_notation_and_review}
Momentum-twistor variables, introduced by Hodges in ref.\ \cite{Hodges:2009hk}, trivialize the two ubiquitous constraints imposed on the external kinematics for all scattering amplitudes: the on-shell condition, $p_a^2\!=\!0$, and momentum conservation, $\sum_ap_a\!=\!0$. By this we mean that {\it generic} momentum twistor variables $Z\!\equiv\!\big(z_1\cdots z_n\big)\!\in\!G(4,n)$, with $z_a\!\in\!\mathbb{C}^4$, {\it always} correspond to a set of momentum-conserving, on-shell external momenta. This makes them especially convenient for use in scattering amplitudes.

In \mbox{section \ref{generalized_unitarity_section}}, we used region-momentum variables $x_a$ to encode the external four-momenta according to $p_a\!\equiv\!x_{a+1}-x_a$. Momentum-twistors are so-called because they represent points in the twistor-space \cite{Penrose:1967wn} of {\it region-momentum} $x$-space:
\vspace{-0.4cm}\eq{\includegraphics[scale=0.875]{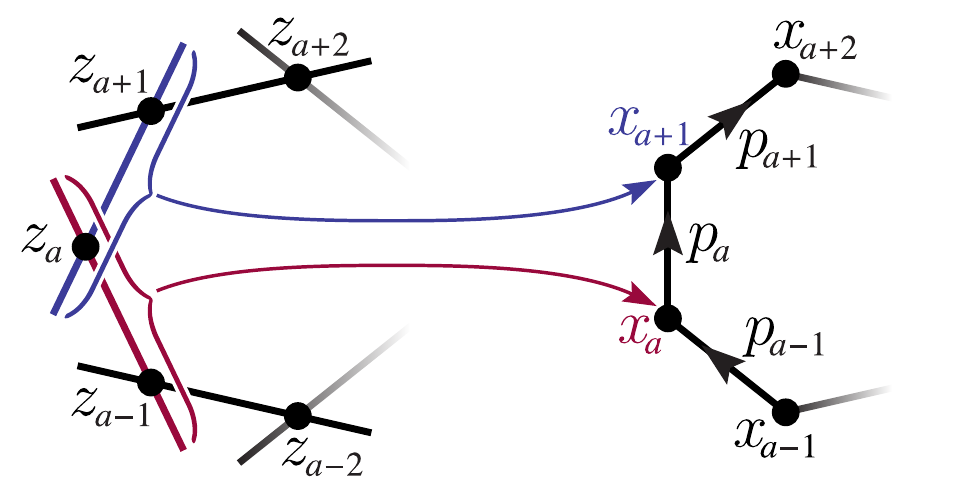}\nonumber\vspace{-0.55cm}}
The $x$-space polygon, whose definition depends on a choice for the cyclic ordering of the external legs, encodes the external momenta in a simple way.
Each line in twistor space---spanned, say by the twistors $(z_{a-1}\,z_{a})$---corresponds to a {\it point} in $x$-space---in this case, the point $x_a$. This is why, for example, integration over a point $x_{\ell}$ in region-momentum space translates to integration over a {\it line} $(\ell)\!\equiv\!(\ell_A\ell_B)$ in momentum-twistor space (see e.g.\ \cite{ArkaniHamed:2010gh,ArkaniHamed:2010kv,ArkaniHamed:2012nw}).

Given a set of momentum-twistors $z_a$---viewed as columns of the \mbox{$(4\!\times\!n)$-matrix} $Z$---it is easy to construct the corresponding set of four-momenta. If we decompose each momentum-twistor $z_a$ according to
\vspace{-0.4cm}\eq{z_a\equiv\big(\lambda_a\,\mu_a\big),\vspace{-0.2cm}\label{components_of_z}} and define a $(2\!\times\!n)$-matrix $\widetilde{\lambda}\!\equiv\!\big(\widetilde\lambda_1\cdots\widetilde\lambda_n\big)\!\subset\!(\lambda^{\perp})$ according to,
\vspace{-0.3cm}\eq{\hspace{-2cm}\widetilde{\lambda}\equiv\mu\!\cdot\!Q\quad\mathrm{with}\quad Q_{ab}\!\equiv\!\frac{\delta_{a-1\,b}\ab{a\,a\pl1}+\delta_{a\,b}\ab{a\pl1\,a\,\mi\,1}+\delta_{a+1\,b}\ab{a\,\mi\,1\,a}}{\ab{a\,\mi\,1\,a}\ab{a\,a\pl1}},\vspace{-0.3cm}\hspace{-2cm}}
where $\ab{a\,b}\!\equiv\!\det\{\lambda_a,\lambda_b\}$, then $\lambda\!\cdot\!\widetilde{\lambda}\!=\!0$ because $Q\!\cdot\!\lambda=0$; as such, we may identify $p_a\!\equiv\!\lambda_a\widetilde{\lambda}_a$, and these (on-shell) four-momenta will {\it automatically} conserve momentum.

Conversely, given four-momenta written in terms of spinor-helicity variables according to $p_a\!\equiv\!\lambda_a\widetilde{\lambda}_a$, momentum twistors $z_a$ can be constructed by joining each $\lambda_a$ as in (\ref{components_of_z}) with a corresponding $\mu_a$ given by
\vspace{-0.4cm}\eq{\mu\equiv \widetilde{Q}\!\cdot\!\widetilde{\lambda}\qquad\mathrm{with}\qquad \widetilde{Q}_{ab}\!\equiv\!\left\{\begin{array}{lr}\ab{b\,a}&\text{if }1\!<\!b\!<\!a\\0&\mathrm{otherwise}\end{array}\right.\,.\vspace{-0.4cm}}

Supersymmetry is encoded by dressing each momentum-twistor $z_a$ with an anti-commuting four-vector $\eta_a$---collected into a $(4\!\times\!n)$-matrix $\eta$ acted upon by the $SU(4)$ $R$-symmetry of $\mathcal{N}\!=\!4$ SYM. If we similarly define $\widetilde{\eta}\!\equiv\!\eta\cdot Q$, then the kinematical data specified by $\{\lambda,\widetilde\lambda,\widetilde\eta\}$ will automatically be {\it super}momentum-conserving.

Dual-conformal transformations in region-momentum space translate to mere $SL(4)$-transformations in momentum-twistor space; hence, dual-conformal invariants are written in terms of simple determinants: $\ab{a\,b\,c\,d}\!\equiv\!\det\{z_a,z_b,z_c,z_d\}$. The simplest dual-{\it super}conformal invariant, however, involves five momentum-twistors, and is given by the familiar $5$-bracket (sometimes called an `$R$-invariant'), \cite{Drummond:2008vq,ArkaniHamed:2012nw}:
\vspace{-0.2cm}\eq{\hspace{-2cm}[a\,b\,c\,d\,e]\equiv\frac{\delta^{1\times4}\big(\eta_a\ab{b\,c\,d\,e}+\eta_b\ab{c\,d\,e\,a}+\eta_c\ab{d\,e\,a\,b}+\eta_d\ab{e\,a\,b\,c}+\eta_e\ab{a\,b\,c\,d}\big)}{\ab{a\,b\,c\,d}\ab{b\,c\,d\,e}\ab{c\,d\,e\,a}\ab{d\,e\,a\,b}\ab{e\,a\,b\,c}}.\vspace{-0.2cm}\hspace{-1.25cm}}

All one-loop leading singularities {\it except the four-masses}\footnote{Recall that the four-mass leading singularities require a prefactor of $\varphi_{{\color{cut1}1},{\color{cut2}2}}$ (see equation (\ref{four_mass_coefficient_1})).} can be written directly as products of $5$-brackets---as evidenced by \mbox{Table \ref{one_loop_leading_singularities_table}} and the fact that BCFW recursion (see \mbox{Appendix \ref{bcfw_loop_recursion_section}}) directly gives tree-amplitudes in terms of products of $5$-brackets. These often involve geometrically-defined, auxiliary points in momentum-twistor space such as ``$(a\,b)\newcap(c\,d\,e)$'' which represents ``$\mathrm{span}\{z_a,z_b\}\newcap\mathrm{span}\{z_c,z_d,z_e\}$''. All such objects are trivially found via {\it Cramer's rule}, which represents the unique identity satisfied by any five generic four-vectors:
\vspace{-0.2cm}\eq{z_a\ab{b\,c\,d\,e}+z_b\ab{c\,d\,e\,a}+z_c\ab{d\,e\,a\,b}+z_d\ab{e\,a\,b\,c}+z_e\ab{a\,b\,c\,d}=0;\vspace{-0.2cm}}
from this, it is easy to see that
\vspace{-0.2cm}\eq{\hspace{-3cm}\begin{array}{@{}c@{}}(a\,b)\newcap(c\,d\,e)\equiv z_a\ab{b\,c\,d\,e}+z_b\ab{c\,d\,e\,a}=-\left(z_c\ab{d\,e\,a\,b}+z_d\ab{e\,a\,b\,c}+z_e\ab{a\,b\,c\,d}\right)\end{array}.\vspace{-0.2cm}\hspace{-2cm}}
A similar, geometrically-defined object which appears in the BCFW recursion of one-loop amplitudes (see \mbox{Appendix \ref{bcfw_loop_recursion_section}}) is ``$(a\,b\,c)\newcap(d\,e\,f)$'', which simply represents ``\mbox{$\mathrm{span}\{z_a,z_b,z_c\}\newcap\mathrm{span}\{z_d,z_e,z_f\}$}''---which, in this case, is a (projective) {\it line} in twistor space; for the sake of concreteness, we can always write this explicitly as:\\[-8pt]
\vspace{-0.2cm}\eq{(a\,b\,c)\newcap(d\,e\,f)\equiv(a\,b)\ab{c\,d\,e\,f}+(b\,c)\ab{a\,d\,e\,f}+(c\,a)\ab{b\,d\,e\,f}.\vspace{-0.1cm}}

Finally, we should recall that the Jacobian arising form the change of variables from momentum-space to momentum-twistor space is the full Parke-Taylor MHV super-amplitude \cite{ArkaniHamed:2009vw},
\vspace{-0.2cm}\eq{\frac{\delta^{2\times4}\big(\lambda\!\cdot\!\widetilde{\eta}\big)\delta^{2\times2}\big(\lambda\!\cdot\!\widetilde{\lambda}\big)}{\ab{1\,2}\ab{2\,3}\cdots\ab{n\,1}},\vspace{-0.1cm}\label{parke_taylor}}
 which explains why (when written in momentum-twistors) all MHV amplitudes are simply $\mathcal{A}_n^{(0),0}\!=\!1$, ensuring the dual-conformal symmetry of all amplitudes in planar $\mathcal{N}\!=\!4$ SYM, \cite{Drummond:2008vq}; and so, throughout this paper, $\mathcal{A}_n^{(k),\ell}$ should be understood as the color-ordered, single-trace contribution to the $\ell$-loop integrand for the $n$-point N$^{k}$MHV scattering amplitude divided by (\ref{parke_taylor}), and in units of $(g^2N_C/(16\pi^2))^\ell$.

\section{The BCFW Representation of One-Loop Integrands}\label{bcfw_loop_recursion_section}
As described in ref.\ \cite{ArkaniHamed:2010kv} (see also \cite{ArkaniHamed:2012nw}), all $\ell$-loop integrands for scattering amplitudes in planar $\mathcal{N}\!=\!4$ SYM can be found by the BCFW recursion relations. In terms of on-shell diagrams, the BCFW recursion relations correspond to:
\vspace{-0.2cm}\eq{\hspace{-3cm}\raisebox{-50pt}{\includegraphics[scale=1]{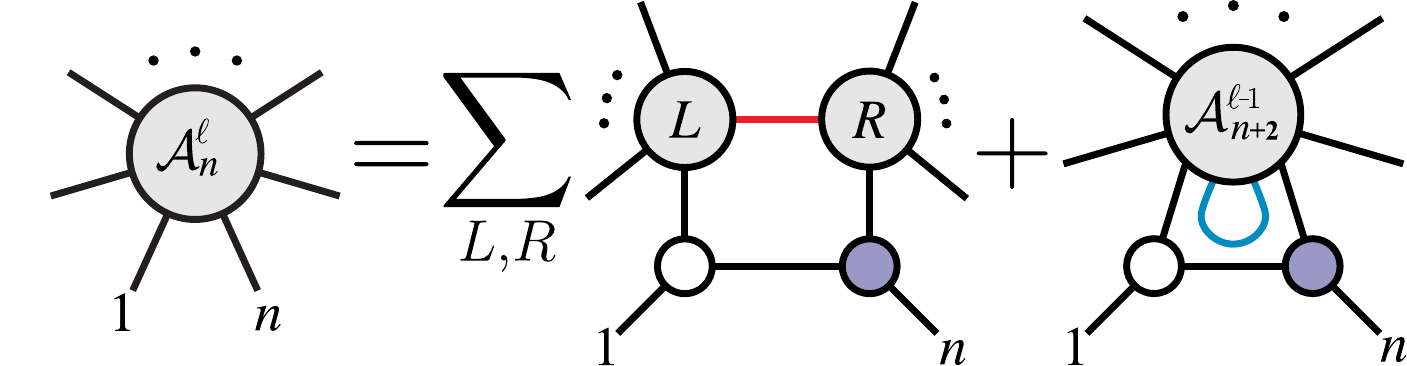}}\hspace{-2cm}\vspace{-0.2cm}}
Being more explicit about the ranges for the terms involved, the recursion becomes,
\vspace{-.2cm}\eq{\mathcal{A}_n^{(k),\ell}\;=\hspace{-.25cm}\sum_{\text{{\scriptsize$\begin{array}{@{}c@{}c@{}c@{$\!$}c@{}c@{}c@{}l}\\[-14pt]n&=&n_L&+&n_R&-&2\\k&=&k_L&+&k_R&+&1\\\ell&=&\ell_L&+&\ell_R\\[-4pt]\end{array}$}}}\!\!\!\!\!\!\mathcal{A}_{n_L}^{(k_L),\ell_L}\!\!\bigotimes_{\mathrm{BCFW}}\!\!\mathcal{A}_{n_R}^{(k_R),\ell_R}+\mathrm{FL}\Big(\mathcal{A}_{n+2}^{(k+1),\ell-1}\Big).\vspace{-.2cm}\label{bcfw_all_loop_recursion}}

Working in momentum-twistor space, the BCFW bridge operation corresponding to the shift $z_n\!\to\!z_n\pl\,\alpha\, z_{n-1}$, for $n_R\!>\!3$, is given by:
\vspace{-.0cm}\eq{\hspace{-.5cm}\mathcal{A}_{n_L}^{(k_L),\ell_L}\!\!\bigotimes_{\mathrm{BCFW}}\!\!\mathcal{A}_{n_R>3}^{(k_R),\ell_R}\equiv\mathcal{A}_{n_L=a}^{(k_L),\ell_L}(1,\ldots,a\mi1,\ahat\,)[1\,a\mi1\,a\,n\mi1\,n]\mathcal{A}_{n_R}^{(k_R),\ell_R}(\,\ahat,a,\ldots,n\mi1,\hat{n}),\vspace{-.2cm}\nonumber}
where $\ahat\equiv(a\,a\mi1)\newcap(n\mi1\,n\,1)$ and $\hat{n}\equiv(n\,n\mi1)\newcap(a\mi1\,a\,1)$; when $n_R\!=\!3$ and $n_L\!=\!n\mi1$, the bridge simply results in,
\vspace{-.0cm}\eq{\mathcal{A}_{n-1}^{(k),\ell_L}\!\!\bigotimes_{\mathrm{BCFW}}\!\!\mathcal{A}_{3}^{(-1),0}\equiv\mathcal{A}_{n-1}^{(k),\ell_L}(1,\ldots,n\mi1).\vspace{-.2cm}\nonumber}
And so, the `bridge' terms of (\ref{bcfw_all_loop_recursion}) are fairly straightforward to compute in momentum-twistor space, and the operations involved are the same regardless of the loop-levels $\ell_L$ and $\ell_R$ of the amplitudes being bridged. (Of course, at tree-level, {\it only} the bridge terms contribute to the recursion; and so the discussion so far suffices to recursively compute all tree-amplitudes, $\mathcal{A}_n^{(k),0}$, in $\mathcal{N}\!=\!4$ SYM.)

More interesting are the ``forward-limit'' contributions, $\mathrm{FL}\big(\mathcal{A}_{n+2}^{(k+1),\ell-1}\big)$. It is easy to see that (\ref{bcfw_all_loop_recursion}) gives rise to $\ell$ levels of nested forward-limits. As described in \cite{ArkaniHamed:2012nw}, determining which terms from the lower-loop amplitude are non-vanishing in the forward limit is generally difficult (even the {\it number} of terms which survive becomes scheme-dependent beyond one-loop). However, for one-loop amplitudes, we only need the forward-limits of trees; and as described in ref.\ \cite{ArkaniHamed:2012nw}, if the tree-amplitudes are obtained by BCFW ,deforming the legs which are to be identified in the forward-limit, then the terms which vanish are {\it precisely} those involving three-particle amplitudes on either side of the bridge. Therefore, the only non-vanishing contributions are:
\eq{\hspace{-.2cm}\begin{array}{rl}\displaystyle\underset{\!\ell_A\to\ell_B}{\mathrm{FL}}\Big(\mathcal{A}_{n+2}^{(k+1),0}(\ell_B,1,\ldots,n,\ell_A)\Big)\hspace{-3.95cm}\\=&\displaystyle\hspace{-.05cm}\sum_{\substack{n_L,n_R\geq4\\k=k_L+k_R}}\hspace{-.25cm}\mathrm{FL}\Big(\mathcal{A}_{n_L=a+1}^{(k_L),0}\hspace{-0.05cm}(\ell_B,1,\ldots,a\mi1,\ahat\,)[\ell_A\,\ell_B\,a\mi1\,a\,n]\mathcal{A}_{n_R}^{(k_R),0}(\,\ahat,a,\ldots,n,\hat{\ell_A})\Big),
\\=&\hspace{-.05cm}\displaystyle\sum_{\substack{n_L,n_R\geq4\\k=k_L+k_R}}\hspace{-.25cm}\mathcal{A}_{n_L=a+1}^{(k_L),0}\hspace{-0.05cm}(\hat{\ell},1,\ldots,a\mi1,\ahat\,)K[(1\,a\mi1\,a);(1\,n\mi1\,n)]\mathcal{A}_{n_R}^{(k_R),0}(\,\ahat,a,\ldots,n\mi1,\hat{n},\hat{\ell}),\\[-0pt]
\end{array}\nonumber}
where $\ahat\!\equiv\!(a\,a\mi1)\newcap(\ell_A\ell_B\,1)$, $\hat{n}\equiv(n\,n\mi1)\newcap(\ell_A\ell_B\,1)$, and $\hat{\ell}\!\equiv\!(\ell_A\ell_B)\newcap(n\mi1\,n\,1)$, and the ``kermit'' $K[(a\,b\,c);(d\,e\,f)]$---written in terms of the line $\ell\!\equiv\!(\ell_A\ell_B)$---is given by:
\vspace{-1cm}\eq{\hspace{-2cm}\begin{array}{rl}\\[-2.5pt]~\\K[(a\,b\,c);(d\,e\,f)]\equiv&\displaystyle -d^4\ell\,\,\frac{\ab{\ell\,(abc)\newcap(def)}^2}{\ab{\ell\,ab}\ab{\ell\,bc}\ab{\ell\,ca}\ab{\ell\,de}\ab{\ell\,ef}\ab{\ell\,fd}},\\[15pt]=&\hspace{-0cm}\displaystyle -d\log\left(\!\frac{\ab{\ell\,ab}}{\ab{\ell\,bc}}\!\right)d\log\left(\!\frac{\ab{\ell\,bc}}{\ab{\ell\,ca}}\!\right)d\log\left(\!\frac{\ab{\ell\,de}}{\ab{\ell\,ef}}\!\right)d\log\left(\!\frac{\ab{\ell\,ef}}{\ab{\ell\,fd}}\!\right).\\[-2.5pt]\end{array}\hspace{-2cm}\vspace{-.0cm}\label{introducing_mister_kermit}}

Putting everything together, the one-loop integrand for any amplitude is:\\[-8pt]
\eq{\hspace{-.0cm}\boxed{\begin{array}{rl}\displaystyle\mathcal{A}_{n}^{(k),1}(1,\ldots,n)\hspace{-2cm}\\[5pt]
&\hspace{-.65cm}=\;\;\hspace{15pt}\mathcal{A}_{n-1}^{(k),1}(1,\ldots,n\mi1)\;\hspace{-4.5cm}
\\[-12pt]\\
&\hspace{-.3cm}+\hspace{-.2cm}\displaystyle\sum_{\text{{\scriptsize$\begin{array}{@{}c@{}c@{}c@{$\!$}c@{}c@{}c@{}l}\\[-14pt]n&=&n_L&+&n_R&-&2\\k&=&k_L&+&k_R&+&1\\1&=&\ell_L&+&\ell_R\\[-4pt]\end{array}$}}}\!\!\!\underset{\;\;\;\;\;\;\;\text{{\normalsize$\begin{array}{c}\\[-23pt]\text{{\footnotesize ~ }}\\\\[-18pt]\text{{\footnotesize$\ahat\equiv(a\,a\mi1)\newcap(n\mi1\,n\,1),\;\hat{n}\equiv(n\,n\mi1)\newcap(a\mi1\,a\,1)$}}\end{array}$}}}{\mathcal{A}_{n_L=a}^{(k_L),\ell_L}(1,\ldots,a\mi1,\ahat\,)[1\,a\mi1\,a\,n\mi1\,n]\mathcal{A}_{n_R}^{(k_R),\ell_R}(\,\ahat,a,\ldots,n\mi1,\hat{n})}
\\[-5pt]\\&\hspace{-.3cm}+\hspace{-.2cm}\displaystyle\sum_{\substack{n=n_L\!+n_R-4\\n_L,n_R\geq4\\k=k_L\!+k_R}}\hspace{-.25cm}\underset{\;\;\;\;\text{{\normalsize$\begin{array}{c}\\[-20pt]\text{{\footnotesize ~ }}\\\\[-23pt]\hspace{-5pt}\text{{\footnotesize$\ahat\equiv(a\,a\mi1)\newcap(\ell_A\ell_B\,1),\;\hat{n}\equiv(n\,n\mi1)\newcap(\ell_A\ell_B\,1),\; \hat{\ell}\equiv(\ell_A\ell_B)\newcap(n\mi1\,n\,1)$}}\end{array}$}}}{\mathcal{A}_{n_L=a+1}^{(k_L),0}\hspace{-0.05cm}(\hat{\ell},1,\ldots,a\mi1,\ahat\,)K[(1\,a\mi1\,a);(1\,n\mi1\,n)]\mathcal{A}_{n_R}^{(k_R),0}(\,\ahat,a,\ldots,n\mi1,\hat{n},\hat{\ell}).}
\end{array}}\nonumber}
Here, the first two lines represent `bridge'
contributions---identical in form to the tree-level
recursion---while the last line represents the forward-limits.
Notice the striking similarity of the roles of the $5$-bracket
$[1\,a\mi1\,a\,n\mi1\,n]$ in the bridge-terms and the `kermit'
$K[(1\,a\,\mi1\,a);(1\,n\mi1\,n)]$ in the forward-limit terms.
Indeed, the forward-limit terms can be understood as unitarity-cuts
which are ``bridged'' by the kermit.

This analysis can of course be continued to higher loop-orders by repeatedly substituting the structure above into the forward-limit contributions appearing in (\ref{bcfw_all_loop_recursion}); this results in higher-loop ``kermits'' which can similarly be understood as `bridging' amplitudes across a unitarity cut. However, as our present work requires only one-loop integrands (and as the the complexity involved in the higher-loop `kermits' is considerable), we will leave a more general discussion to future work.

\newpage
\section{{\sc Mathematica} Implementation of Results}\label{mathematica_implementation_section}
In order to make the results described in this paper most useful to researchers, we have prepared a {\sc Mathematica} package called `{\tt loop\uscore amplitudes}' which implements all of our results. In addition to providing fast numerical evaluation of loop amplitudes and ratio functions, the {\tt loop\uscore amplitudes} package also serves as a reliable reference for the many results tabulated above (as any transcription error would obstruct numerical consistency checks).

The package together with a notebook illustrating much of its functionality are included with the submission files for this paper on the {\tt arXiv}, which can be obtained as follows. From the abstract page for this paper on the {\tt arXiv}, look for the ``download'' options in the upper-right corner of the page, follow the link to ``other formats'' (below the option for ``PDF''), and download the ``source'' files for the submission. The source will contain\footnote{Occasionally, the ``source'' file downloaded from the {\tt arXiv} is saved without any extension; this can be ameliorated by manually appending ``{\tt .tar.gz}'' to the name of the downloaded file.} the primary package ({\tt loop\rule[-1.05pt]{7.5pt}{.75pt}amplitudes.m}), together with a notebook ({\tt loop\rule[-1.05pt]{7.5pt}{.75pt}amplitudes\rule[-1.05pt]{7.5pt}{.75pt}demo.nb}) which includes many detailed examples of the package's functionality.

Upon obtaining the source files, one should open and evaluate the {\sc Mathematica} notebook `{\tt loop\rule[-1.05pt]{7.5pt}{.75pt}amplitudes\rule[-1.05pt]{7.5pt}{.75pt}demo.nb}'; in addition to walking the user through many example computations, this notebook will copy the file  {\tt loop\rule[-1.05pt]{7.5pt}{.75pt}amplitudes.m} to the user's {\tt ApplicationDirectory[]}; this will make the package available to run in any future notebook via the simple command ``{\tt <<loop\uscore amplitudes.m}'':\\[-15pt]

\mathematica{.8}{\raisebox{-2pt}{{\tt<<loop\rule[-1.05pt]{7.5pt}{.75pt}amplitudes.m}}}{\raisebox{-200pt}{\includegraphics[scale=.875]{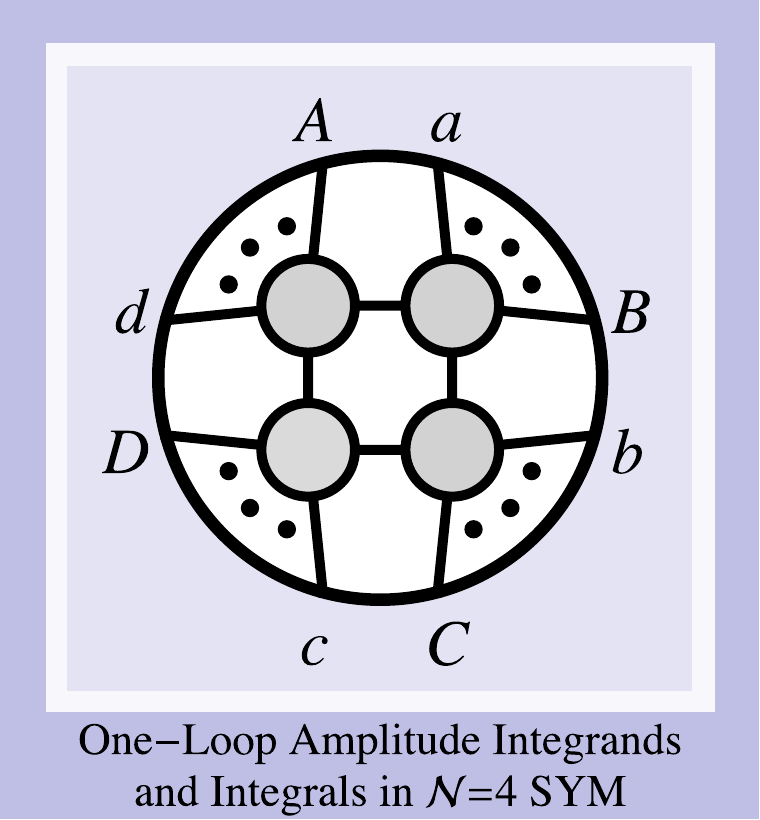}}}\\[-20pt]

\newpage
\subsection{Glossary of Functions Defined By the {\sc Mathematica} Package}

\noindent{\large{\bf Abstract Representations of Box Expansions}}

\defntb{boxCoefficient}{\vardef{n},\vardef{k},\vardefo{cut}{\tt 0}}{\vardefms{abcd}}{returns the scalar box coefficient $f_{\var{abcd}}$ for the amplitude $\mathcal{A}_{\var{n}}^{(\var{k}),1}$ if \var{cut} is {\tt 0} (its default); alternatively, it will give ${\color{cut1}f^{1}_{\var{abcd}}}$ or ${\color{cut2}f^{2}_{\var{abcd}}}$ if \var{cut} is set to {\tt 1} or {\tt 2}, respectively.
}

\defn{chiralBoxExpansion}{\vardef{n},\vardef{k}}{returns an abstract (symbolic) list of contributions to the one-loop amplitude $\mathcal{A}_{\var{n}}^{(\var{k}),1}$ in terms of on-shell diagrams and {\it chiral} boxes.
}

\defn{chiralBoxRatioExpansion}{\vardef{n},\vardef{k}}{returns an abstract (symbolic) list of contributions to the one-loop ratio function $\mathcal{R}_{\var{n}}^{(\var{k}),1}\!=\!\mathcal{A}_{\var{n}}^{(\var{k}),1}\mi\,\mathcal{A}_{\var{n}}^{(\var{k}),0}\mathcal{A}_{\var{n}}^{(0),1}$ written in terms of on-shell diagrams and {\it chiral} boxes.
}

\defn{onlyCyclicSeeds}{\vardef{expression}}{returns the parts of a symbolic \var{expression} for a loop integrand or integral which generate the entire \var{expression} upon summing over cyclic classes.
}

\defn{scalarBoxExpansion}{\vardef{n},\vardef{k}}{returns an abstract (symbolic) list of contributions to the one-loop amplitude $\mathcal{A}_{\var{n}}^{(\var{k}),1}$ in terms of on-shell diagrams and {\it scalar} boxes.
}

\defn{scalarBoxRatioExpansion}{\vardef{n},\vardef{k}}{returns an abstract (symbolic) list of contributions to the one-loop ratio function $\mathcal{R}_{\var{n}}^{(\var{k}),1}\!=\!\mathcal{A}_{\var{n}}^{(\var{k}),1}\mi\,\mathcal{A}_{\var{n}}^{(\var{k}),0}\mathcal{A}_{\var{n}}^{(0),1}$ written in terms of on-shell diagrams and {\it scalar} boxes.
}

~\\\noindent{\large{\bf (Internal) Symbols for Objects Used by {\tt loop\uscore amplitudes} }}

\defntb{chiralBox}{\vardef{cut}}{\vardef{legList}}{an {\it undefined} object that is used internally by the package to represent the chiral-box $\widetilde{I}^{\var{cut}}$---for \var{cut} either {\color{cut1}{\tt 1}} or {\color{cut2}{\tt 2}}---involving the corners whose external legs are specified by \var{legList}. To be clear, given \var{legList} expressed in the form \var{legList}$\,\equiv\!{\tt \{\{}\var{a},\ldots,\var{B}{\tt\}},\ldots,{\tt \{}\var{d},\ldots,\var{D}{\tt \}\}}$, \fun{chiralBox} would (abstractly) represent the chiral box $\widetilde{I}^{\var{cut}}_{\var{abcd}}$ (see \mbox{Table \ref{chiral_box_integrands_and_integrals_table}}).
}

\defntb{onShellGraph}{\vardef{cut}}{\vardef{legList},\vardef{kList}}{an {\it undefined} object that is used internally by the package to represent the on-shell diagram involving the corner amplitudes of type N$^{\var{k}}$MHV with $\var{k}\!\in\!\var{kList}$, and involving the legs specified by \var{legList}. To be clear, given \var{legList}$\,=\!{\tt \{\{}\var{a},\ldots,\var{B}{\tt\}},\ldots,{\tt \{}\var{d},\ldots,\var{D}{\tt \}\}}$ and \var{kList}$\,=\!{\tt\{}\var{k_a},\ldots,\var{k_d}{\tt \}}$, then \fun{onShellGraph} would (abstractly) represent the on-shell diagram whose corner-amplitudes are given by,
\vspace{-0.2cm}\eq{\begin{array}{rccl}{\tt \{}&\mathcal{A}_{n_a}^{(\var{k_a}),0}(\alpha_{{\color{cut1}1},{\color{cut2}2}},\var{a},\ldots,\var{B},\beta_{{\color{cut1}1},{\color{cut2}2}}),&\mathcal{A}_{n_b}^{(\var{k_b}),0}(\beta_{{\color{cut1}1},{\color{cut2}2}},\var{b},\ldots,\var{C},\gamma_{{\color{cut1}1},{\color{cut2}2}}),\\
&\mathcal{A}_{n_c}^{(\var{k_c}),0}(\gamma_{{\color{cut1}1},{\color{cut2}2}},\var{c},\ldots,\var{D},\delta_{{\color{cut1}1},{\color{cut2}2}}),&\mathcal{A}_{n_d}^{(\var{k_d}),0}(\delta_{{\color{cut1}1},{\color{cut2}2}},\var{d},\ldots,\var{A},\alpha_{{\color{cut1}1},{\color{cut2}2}})&{\tt \}},
\end{array}\vspace{-0.0cm}}
for the quad-cut solution ${\color{cut1}\ell_1}$ or ${\color{cut2}\ell_2}$ depending on whether \var{cut} is {\color{cut1}{\tt 1}} or {\color{cut2}{\tt 2}}, respectively.
}

\defn{R}{\vardefms{abcde}}{represents an {\it undefined} object that is used internally by the package to represent the $5$-bracket involving twistors given by the sequence \var{abcde}.
}

\defn{scalarBox}{\vardef{legList}}{an {\it undefined} object that is used internally by the package to represent the scalar box $I_{\var{abcd}}$ whose corners involve the external legs specified by \var{legList}, given in the form: \var{legList}$\,\equiv\!{\tt \{\{}\var{a},\ldots,\var{B}{\tt\}},\ldots,{\tt \{}\var{d},\ldots,\var{D}{\tt \}\}}$.
}

~\\\noindent{\large{\bf Analytic Expressions for Objects Involved in Box Expansions}}

\defn{boxRs}{\vardef{legList}}{returns the list $\{{\color{cut1}R_1},{\color{cut2}R_2}\}$ of the $5$-bracket prefactors---for $\{{\color{cut1}\ell_1},{\color{cut2}\ell_2}\}$, respectively---(obtained as an on-shell graph where all corners other than those involving $\mathcal{A}_3^{(-1)}$ have been replaced by MHV amplitudes) which multiplies the corner amplitudes to result in a given one-loop leading singularity. These are listed in detail in \mbox{Table \ref{one_loop_leading_singularities_table}}.
}

\defn{chiralBoxIntegral}{\vardef{legList}}{returns the (manifestly finite) integrated form of the chiral box whose corners involve the legs \var{legList}. (Recall that both chiral integrands $\{{\color{cut1}\tilde{\mathcal{I}}^1},{\color{cut2}\tilde{\mathcal{I}}^2}\}$---being parity conjugates of one another---integrate to $I^{{\color{cut1}1},{\color{cut2}2}}$ because loop integration is always performed on a parity-invariant contour.) These integrals are listed in \mbox{Table \ref{chiral_box_integrands_and_integrals_table}}.
}

\defn{chiralBoxIntegrands}{\vardef{legList}}{returns a list of the of chiral integrands $\{{\color{cut1}\tilde{\mathcal{I}}^1},{\color{cut2}\tilde{\mathcal{I}}^2}\}$ for the box whose corners involve the legs specified by \var{legList}; the numerators of these integrands are listed in \mbox{Table \ref{chiral_box_integrands_and_integrals_table}}.
}

\defn{quadCuts}{\vardef{legList}}{for a box whose corners are given by the legs specified by \var{legList}, \fun{quadCuts} returns $\{\{{\color{cut1}\alpha_1},\ldots,{\color{cut1}\delta_1}\},\{{\color{cut2}\alpha_2},\ldots,{\color{cut2}\delta_2}\}\}$, specifying the points along $(Aa),\ldots,(Dd)$ which lie on the quad-cuts ${\color{cut1}\ell_1}$ and ${\color{cut2}\ell_2}$, respectively; see \mbox{Table \ref{quad_cuts_table}}.
}

\defn{scalarBoxIntegral}{\vardef{legList},\vardefo{keepEpsilonsQ}{\tt True}}{returns the integrated form of the scalar box whose corners are specified by \var{legList} using the DCI regulator described in \mbox{section \ref{DCI_regulator_subsection}} (see \mbox{Table \ref{dci_regulated_box_integrals}}). If the optional second argument of \var{keepEpsilonsQ} is set to {\tt False}, then all the divergences---proportional to $\log(\epsilon)$---will be excluded, returning only $I^{\mathrm{fin}}$.
}

\defn{scalarBoxIntegrand}{\vardef{legList}}{returns the canonically-normalized scalar box integrand for the box whose corners are specified by \var{legList}.
}

~\\\noindent{\large{\bf Replacing Abstract Objects with Analytic Formulae}}

\defntb{fromRform}{\vardef{n}}{\vardef{expression}}{converts any $5$-brackets in \var{expression} (encoded abstractly as \fun{R}[\var{a},\var{b},\var{c},\var{d},\var{e}]) into explicit superfunctions encoded by pairs $\{f,C\}$.
}

\defn{integrand}{\vardef{expression}}{converts any abstract boxes (\fun{scalarBox} or \fun{chiralBox}) occurring in \var{expression} to explicit integrands involving the loop-momentum $(\ell)$---represented by the line $(\ell_A\ell_B)$ in momentum-twistor space.
}

\defn{integrate}{\vardef{expression}}{converts any abstract boxes (\fun{scalarBox} or \fun{chiralBox}) occurring in \var{expression} to explicit, integrated expressions.
}

\defn{rForm}{\vardef{expression}}{converts any abstract on-shell graph objects (\fun{onShellGraph}) occurring in \var{expression} to explicit formulae written in terms of $5$-brackets.
}

~\\\noindent{\large{\bf BCFW-Recursion of Tree-Amplitudes and Loop Integrands}}

\defn{bcfwLoopIntegrand}{\vardef{n},\vardef{k}}{returns the `superfunction' form of \fun{rAmp}[$\var{n},\var{k},1$]---that is, not written in terms of $5$-brackets, but in terms of pairs $\{f,C\}$ which encode a superfunction $f\!\times\!\delta^{k\times4}\big(C\!\cdot\!\eta)$. \fun{bcfwLoopIntegrand}[\var{n},\var{k}] is equivalent to \fun{fromRform}[\var{n}][\fun{rAmp}[\var{n},\var{k},1]].
}

\defn{rAmp}{\vardef{n},\vardef{k},\vardefo{$\ell$}0}{returns the tree (\var{$\ell$}$=${\tt0}) or one-loop (\var{$\ell$}$=${\tt1}) amplitude $\mathcal{A}_{\var{n}}^{(\var{k}),\var{\ell}}$ as obtained via BCFW recursion---written in terms of $5$-brackets and `kermits' as described in \mbox{Appendix \ref{bcfw_loop_recursion_section}}.
}

\defn{treeAmp}{\vardef{n},\vardef{k}}{returns the `superfunction' form of \fun{rAmp}[$\var{n},\var{k},0$]---that is, not written in terms of $5$-brackets, but in terms of pairs $\{f,C\}$ which encode a superfunction $f\!\times\!\delta^{k\times4}\big(C\cdot\eta)$. \fun{treeAmp}[\var{n},\var{k}] is equivalent to \fun{fromRform}[\var{n}][\fun{rAmp}[\var{n},\var{k}]].
}

~\\\noindent{\large{\bf Box Expansions of Loop Amplitudes and Ratio Functions}}

\defn{chiralLoopAmp}{\vardef{n},\vardef{k}}{returns a list of superfunctions contributing to the one-loop amplitude $\int\!\!d^4\ell\,\mathcal{A}_{\var{n}}^{(\var{k}),1}$ obtained using {\it chiral} boxes. \fun{chiralLoopAmp}[\var{n},\var{k}] is equivalent to \fun{analyticIntegral}[\fun{chiralBoxExpansion}[\var{n},\var{k}]].
}

\defn{chiralLoopIntegrand}{\vardef{n},\vardef{k}}{returns a list of superfunctions contributing to the one-loop amplitude integrand $\mathcal{A}_{\var{n}}^{(\var{k}),1}$ obtained using {\it chiral} boxes. \fun{chiralLoopIntegrand}[\var{n},\var{k}] is equivalent to \fun{analyticIntegrand}[\fun{chiralBoxExpansion}[\var{n},\var{k}]].
}

\defn{chiralRatioIntegral}{\vardef{n},\vardef{k}}{returns a list of superfunctions contributing to the one-loop ratio function $\int\!\!d^4\ell\,\mathcal{R}_{\var{n}}^{(\var{k}),1}$ obtained using {\it chiral} boxes. \fun{chiralRatioIntegral}[\var{n},\var{k}] is equivalent to \fun{analyticIntegral}[\fun{chiralBoxRatioExpansion}[\var{n},\var{k}]].
}

\defn{loopAmp}{\vardef{n},\vardef{k}}{returns a list of superfunctions contributing to the one-loop amplitude $\int\!\!d^4\ell\,\mathcal{A}_{\var{n}}^{(\var{k}),1}$ obtained using {\it scalar} boxes. \fun{loopAmp}[\var{n},\var{k}] is equivalent to \fun{analyticIntegral}[\fun{scalarBoxExpansion}[\var{n},\var{k}]].
}

\defn{ratioIntegral}{\vardef{n},\vardef{k}}{returns a list of superfunctions contributing to the one-loop ratio function $\int\!\!d^4\ell\,\mathcal{R}_{\var{n}}^{(\var{k}),1}$ obtained using {\it scalar} boxes. \fun{ratioIntegral}[\var{n},\var{k}] is equivalent to \fun{analyticIntegral}[\fun{scalarBoxRatioExpansion}[\var{n},\var{k}]].
}

\newpage
~\\\noindent{\large{\bf Kinematical Specification and Numerical Evaluation}}

\defn{evaluate}{\vardef{expression}}{numerically evaluates all superfunctions occurring in \var{expression} for the kinematical data defined by the global, $(n\!\times\!4)$ matrix {\tt Zs}. If \var{expression} involves an auxiliary line $(X)$ or a loop-variable $(\ell)\!\equiv\!(\ell_A\ell_B)$, these are taken to be given by the last four entries of the global matrix {\tt Zs}---that is, \mbox{{\tt Zs}$\equiv\!\{z_1,\ldots,z_n,z_{X_1},z_{X_2},z_{\ell_A},z_{\ell_B}\}.$}
}

\defn{exampleTwistors}{\vardef{n}}{it is sometimes convenient to evaluate analytic expressions using explicit kinematical data; under such circumstances, there are some conveniences afforded by using ``well-chosen''  kinematical data.\\[-22pt]

\ind Reasons for preferring one choice over another include: having all Lorentz invariants be integer-valued and relatively small; having all dual-conformal cross-ratios {\it positive} (so as to avoid branch-ambiguities when evaluating the polylogarithms that arise in scattering amplitudes at loop-level); and possibly to have all Lorentz-invariants be distinct (either to help reconstruct an analytic expression or to avoid `accidental' cancelations). Of these, the following momentum-twistors meet the first two desires spectacularly:
\vspace{-.2cm}\eq{{\tt Zs}\equiv\left(\begin{array}{@{}cccccc@{}}1&1&1&1&\cdots&\binom{n}{0}\\2&3&4&5&\cdots&\binom{n+1}{1}\\3&6&10&15&\cdots&\binom{n+2}{2}\\4&10&20&35&\cdots&\binom{n+3}{3}\end{array}\right).\vspace{-.2cm}}

\ind The function \fun{exampleTwistros}{\tt [16]} is evaluated when the {\tt loop\rule[-1.05pt]{7.5pt}{.75pt}amplitudes} package is first loaded, allowing amplitudes involving as many as $16$ particles to be evaluated without specific initialization.
}

\defn{randomPositiveZs}{\vardef{n}}{picks random kinematical data for which all cross-ratios are positive (the data {\tt Zs} is {\it positive} when viewed as a four-plane: {\tt Zs}$\in\!G_+(4,n)$).
}

\defn{setupUsingSpinors}{\vardef{lambdaList},\vardef{lambdaBarList}}{sets up the global variables {\tt Ls} and {\tt Lbs} for $\lambda$ and $\widetilde{\lambda}$, respectively, and defines the global $(n\!\times\!4)$ matrix {\tt Zs} for momentum-twistors for use in numerical evaluation.
}

\defn{setupUsingTwistors}{\vardef{twistorList}}{sets up the global $(n\!\times\!4)$ matrix {\tt Zs} encoding the momentum-twistor kinematical data, and defines the auxiliary variables {\tt Ls} and {\tt Lbs} for $\lambda$ and $\widetilde{\lambda}$, respectively.
}

\defnNA{showTwistors}{}{returns a formatted table illustrating the kinematical data---as encoded by the currently used ones for evaluation by \fun{evaluate}{\tt[]}.
}

\defntb{superComponent}{\vardefms{component}}{\vardef{superFunction}}{in the {\tt loop\rule[-1.05pt]{7.5pt}{.75pt}amplitudes} package, a \var{superFunction} is always represented by a pair $\{f,C\}$---an {\it ordinary} function $f(1,\ldots,n)$ of the kinematical variables times a {\it fermionic} $\delta$-function of the form,\\[-5pt]
\vspace{-.2cm}\eq{\delta^{k\times4}\big(C\!\cdot\!{\eta}\big)\equiv\prod_{I=1}^4\left\{\bigoplus_{a_1<\!\cdots<a_k}\!\!(a_1\!\cdots a_k)\,\,{\eta}_{a_1}^I\!\!\cdots{\eta}_{a_k}^{I}\right\},\vspace{-.2cm}}
where $C$ is an $(n\times k)$-matrix of ordinary functions, and for each $a=1,\ldots,n$, ${\eta}_a$ is a fermionic (anti-commuting) variable. To be clear, we consider each particle as a Grassmann coherent state \cite{ArkaniHamed:2008gz} of the form,
\vspace{-.2cm}\eq{\left|a \right> \equiv \left|a\right>_{\{\}} +  \eta_a^I  \left|a\right>_{\{I\}} + \frac{1}{2!}  \eta_a^I  \eta_a^J \left|a \right>_{\{I,J\}} + \frac{1}{3!}  \eta_a^I  \eta_a^J  \eta_a^K \left|a\right>_{\{I,J,K\}} +  \eta_a^1  \eta_a^2  \eta_a^3  \eta_a^4 \left|a\right>_{\{1,2,3,4\}}; \nonumber\vspace{-.2cm}}
and if we use \var{$r_a$} to denote the $R$-charge of the $a^{\mathrm{th}}$ particle according to,
\eq{\begin{array}{|l|l|l@{\hspace{1cm}}|l|}
\hline \text{field }&\text{helicity}&R\text{-charge} (\text{\var{$r_a$}})&\text{short-hand for \var{$r_a$}}\\\hline
|a\rangle_{\{\}}&\;\;+1&{\tt \{\}}& {\tt p}\\
|a\rangle_{\{I\}}&\;\;+\frac{1}{2}&{\tt \{I\}}&{\tt p/2}(\Leftrightarrow{\tt \{4\}})\\
|a\rangle_{\{I,J\}}&\;\;\phantom{+}0&{\tt \{I,J\}}&\text{---}\\
|a\rangle_{\{I,J,K\}}&\;\;-\frac{1}{2}&{\tt \{I,J,K\}}&{\tt m/2}(\Leftrightarrow{\tt \{1,2,3\}})\\
|a\rangle_{\{1,2,3,4\}}&\;\;-1&{\tt \{1,2,3,4\}}&{\tt m}\\\hline
\end{array}\nonumber}
then \fun{superComponent}[\var{$r_1$},\ldots,\var{$r_n$}][\var{superFunction}] returns the {\it component} function of \var{superFunction} proportional to,
\vspace{-.4cm}\eq{\prod_{a=1}^{n}\prod_{I\in r_a}{\eta}_a^{I}.\vspace{-.2cm}}
---that is, the component-function involving the states $|1\rangle_{\text{\var{$r_1$}}}\cdots |n\rangle_{\text{\var{$r_n$}}}$.}

~\\\noindent{\large{\bf General Purpose and Functions and \AE sthetical Presentation}}

\defn{nice}{\vardef{expression}}{formats \var{expression} to display `nicely' by making replacements such as {\tt ab[x$\cdots$y]}$\mapsto\!\ab{x\cdots y}$, $\alpha${\tt [1]}$\mapsto\!\alpha_1$, etc., by writing any level-zero matrices in {\tt MatrixForm}, and by drawing figures to represent abstract objects given by objects such as \fun{onShellGraph} or \fun{scalarBox}.
}

\defn{niceTime}{\vardef{timeInSeconds}}{converts a time measured in seconds \var{timeInSeconds}, to human-readable form. For example,\\[3pt]
\mathematica{0.9}{niceTime[299\,792\,458]\hspace{14cm}$~$
niceTime[3.1415926535]}{{\tt 9 years, 182 days}\hspace{14cm}$~$ {\tt 3 seconds, 141 ms}}
}

\defn{random}{\vardef{objectList}}{returns a random element from (the first level of) \var{objectList}.
}

\defn{timed}{\vardef{expression}}{evaluates \var{expression} and prints a message regarding the time required for evaluation.
}

~\newpage

\newpage
\providecommand{\href}[2]{#2}\begingroup\raggedright\endgroup

~\newpage\thispagestyle{empty}
~\newpage\thispagestyle{empty}
~\vspace{\fill}
\eq{\hspace{-5.0cm}\raisebox{-145pt}{\includegraphics[scale=1.4]{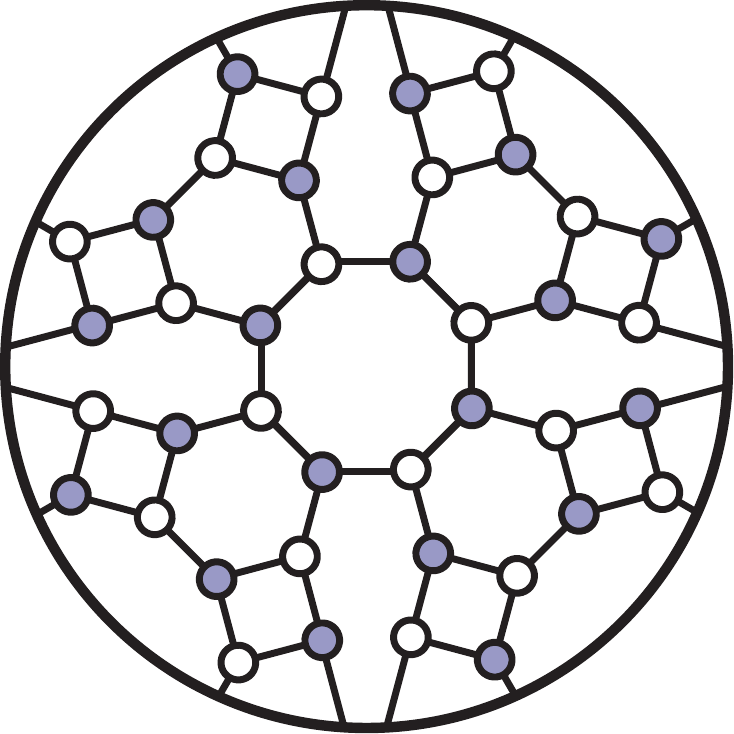}}\hspace{-5cm}\nonumber}
~\vspace{\fill}

\end{document}